# Geometric Partitioning and Ordering Strategies for Task Mapping on Parallel Computers

Mehmet Deveci, Karen D. Devine, Kevin Pedretti,
Mark A. Taylor, Sivasankaran Rajamanickam, (*Member, IEEE*),
and Ümit V. Çatalyürek (*Fellow, IEEE*)

**Abstract**—We present a new method for mapping applications' MPI tasks to cores of a parallel computer such that applications' communication time is reduced. We address the case of sparse node allocation, where the nodes assigned to a job are not necessarily located in a contiguous block nor within close proximity to each other in the network, although our methods generalize to contiguous allocations as well. The goal is to assign tasks to cores so that interdependent tasks are performed by "nearby" cores, thus lowering the distance messages must travel, the amount of congestion in the network, and the overall cost of communication. Our new method applies a geometric partitioning algorithm to both the tasks and the processors, and assigns task parts to the corresponding processor parts. We also present a number of algorithmic optimizations that exploit specific features of the network or application. We show that, for the structured finite difference mini-application MiniGhost, our mapping methods reduced communication time up to 75% relative to MiniGhost's default mapping on 128K cores of a Cray XK7 with sparse allocation. For the atmospheric modeling code E3SM/HOMME, our methods reduced communication time up to 31% on 32K cores of an IBM BlueGene/Q with contiguous allocation.

## 1 INTRODUCTION

Task mapping — the assignment of a parallel application's tasks to the processors of a parallel computer — is increasingly important as the number of computing units in new supercomputers grows from $O(100K)$ to $O(1M)$ and beyond. With large-diameter networks in these supercomputers and many users submitting jobs of various sizes, processor allocations (sets of processors assigned by a job scheduler to parallel jobs) can become sparse and be spread far across the entire network. As a result, communication messages can travel long routes in the network and network links may become congested, which makes maintaining scalability in large-scale machines difficult. These effects can be lessened through the use of topology-aware task mapping. Recent experiments have shown that task mapping can significantly impact performance of parallel applications (e.g., [1], [4], [10], [15], [19], [30]); one application exhibited a 1.64X speedup due to improved mapping [24]

Most parallel scientific computing applications ignore the details of the underlying computer network; rather, they use generic task-to-rank mapping. Such generic mappings become problematic when the rank ordering scheme of the underlying network and task ordering scheme of the application do not match. For example, IBM's BlueGene/Q has a 5D torus network with dimensions A, B, C, D, and E. The default rank ordering first places consecutive ranks within a node, and then along dimensions E, D, C, B and

A. In this architecture, contrary to what one would believe, an application that assigns nearby MPI ranks (say, using graph partitioning) to tasks that communicate with each other will likely face a problem. Such applications will have most of their communication within the nodes first and then along E or D. As a result, the links in A or B will not be fully utilized. On the other hand, Cray Gemini interconnection networks use a space filling curve (SFC) algorithm to order the ranks within the set of allocated nodes. As a result, consecutive MPI ranks are placed first within a node, then in nodes that are close within SFC. The application in the previous example will have a good utilization of the network with these ordering schemes. However, an application using a 3D domain and orders its tasks along $x$, $y$ and $z$ spatial dimensions is likely to have communication imbalances. Such applications will perform most of their $x$-dimension communication within a node, but their communication along $y$ and $z$ are likely to happen with other nodes. Both of these examples demonstrate that an effective mapping of tasks to processors should consider both the communication pattern of the tasks and the physical network topology to reduce application communication cost.

We propose a new task mapping strategy that uses geometric data to represent application tasks and compute resources. We define metrics based on this geometric data to represent the cost of communication between tasks, and use these metrics to evaluate and select effective mappings.

Much research has focused on mapping tasks to block-based allocations, such as those on IBM's BlueGene systems (e.g., [3], [10], [24], [42]). Recently, the task mapping methods for sparse allocations on 3D torus [19], [21], [28], dragonfly [40], slim fly [25], and fat tree [34] networks have been explored. We address both contiguous allocations and non-

- *Deveci, Rajamanickam, Pedretti, Taylor, and Devine are with the Center for Computing Research at Sandia National Laboratories, Albuquerque, NM. E-mail: {mndevec,kddevin,ktpedre,mataylo,srajama}@sandia.gov*
- *Çatalyürek is with the School of Computational Science and Engineering at Georgia Institute of Technology. E-mail: umit@gatech.edu*

contiguous (i.e., sparse) allocations, in which nodes from any portion of the machine can be allocated to a job without regard to the allocation's shape or locality. Such allocations are used in many parallel systems (e.g., Cray, clusters). Mapping strategies developed for general allocations can be used directly for more restrictive block allocations.

Most previous non-contiguous approaches have represented tasks' communication patterns and network topologies as graphs; graph algorithms were then applied to find good mappings. Finding optimal topology mappings is NP-Complete [28], so heuristics are often used to reduce complexity (e.g., [7], [11], [12], [16], [17], [19], [22], [31], [32], [35]). We, instead, use inexpensive geometric partitioning to reorder tasks and processors based on their geometric locality, and use the reordering to map tasks that are "close" to each other geometrically to processors that are "close" to each other in the mesh or torus. Initial experimentation with geometric approaches proved promising [21], [33]. This work improves the previous geometric strategies with a new ordering method and ideas for optimizing mapping on various networks. We demonstrate the benefit of these ideas in applications on two different architectures with heterogenous network links.

General-purpose, open-source graph-based mapping algorithms are available. The LibTopoMap library [28] requires as input a task-communication graph describing the amount of communication between tasks, as well as static files describing the network topology. It uses the ParMETIS graph partitioner [29] to divide tasks into $n$ parts, where $n$ is the number of nodes in the allocation, and then applies a graph algorithm (Greedy, Recursive Bisection, Reverse Cuthill-McKee) to map the parts to nodes. The JOSTLE [41] and Scotch [38] libraries combine mapping with load balancing by using recursive bisection of both network-topology and application-data graphs to partition data and map the resulting parts to processors. Like these libraries, our approach is designed for general-purpose use in applications and is available in the Zoltan2 [13] library.

The main contributions of this paper follow.

• We present a new geometric algorithm for task mapping in both contiguous and non-contiguous processor allocations (Section 4). We show that it is possible to use the same foundational algorithm for various applications and various architectures by performing different optimizations and transformations to the input coordinates to account for architecture and application specific characteristics, such as heterogeneity in the performance of network links.

• We present a new ordering scheme and show its effectiveness for task mapping.

• We demonstrate our algorithm in two applications on up to 128K cores, and assess the ability of our mapping to reduce application communication cost (Section 5).

• We compare our geometric mapping to the default mapping used in applications and to application-specific optimizations; we show that our geometric mapping reduces both communication time and communication metrics for the target applications relative to other methods (Section 5).

## 2 TARGETED COMPUTING ENVIRONMENTS

Geometric task mapping requires that the coordinates of the nodes provide information about the network topology.

Such methods can be applied to mesh- or torus-based interconnection networks. However, to apply the geometric mapping methods to other networks such as Fat-Tree or DragonFly networks, one needs to pre-process the node coordinates to approximate the network topology. Mesh- or torus-based networks are common in parallel computers; for example, Cray's XT, XE, XK computers and IBM's BlueGene computers have torus-based networks. Our target environments include Cray XE6 and XK7 machines (e.g., Titan at Oak Ridge National Laboratory), and IBM BlueGene/Q machines (e.g., Mira at Argonne National Laboratory).

In the Cray XE6 and XK7 3D torus, each Gemini router has connections to six neighboring routers, two each in the $x$, $y$ and $z$ dimensions. Messages' routes between nodes can be represented as a path of "hops" along network links in $x$, $y$ and $z$. Differences in bandwidth can exist depending on the physical connections (e.g., backplane, mezzanine, cable) used. For example, X cables are uniform and have bandwidth of 75 GB/s. Two types of links are used in $y$: Y cables and Y mezzanine traces. These cable and mezzanine links have 37.5 GB/s and 75 GB/s bandwidth, respectively. Similarly, Z links consist of Z cables with 75 GB/s and Z backplane traces with 120 GB/s bandwidths. IBM's Blue-Gene/Q (BG/Q) machines have 5D torus networks; the links have uniform bandwidth along all dimensions.

In mesh- and torus-based systems, "coordinates" of the routers within the network are often available via calls to a system library. A router with 3D coordinates $(i, j, k)$ can communicate with a router with coordinates $(i+1, j+1, k+1)$ via a three-hop path, with one hop in each dimension ($x$, $y$, $z$). A torus provides wrap-around, so that messages take the shortest path (i.e., proceed in the positive or negative direction) along each dimension. In the Cray XK7, these coordinates are available from the Resiliency Communication Agent (RCA) tool (`rca_get_meshcoord`). On the IBM BlueGene/Q, they can be obtained using TopoMgr [8]. Each MPI process can obtain the coordinates of the router to which its compute node is attached. Our task-mapping methods then use $SPL$ between routers within the network as an approximation of communication cost between MPI processes in the corresponding nodes.

Each router in a mesh/torus network is typically connected to one or more multicore compute nodes. In the Cray XK7, for example, each router connects two nodes. The Cray platform Titan has 16 cores per node. Parallel applications can use from one MPI process per node (with threaded parallelism within the node) to one MPI process per core (with shared-memory message passing within the node). In the latter case, co-locating interdependent MPI processes within a node reduces communication over the network, and, thus, reduces execution time. Our task-mapping experiments address this case, but they can be applied without loss of generality to the multithreaded case as well.

One difference between the Cray and IBM systems is the way they allocate compute nodes to jobs. In IBM systems, jobs are given a contiguous block of nodes within the network; each dimension of this block is a power of two. In contrast, Cray systems allocate non-contiguous sets of nodes of any size requested by the user. Available nodes are selected according to a space-filling curve algorithm in the ALPS scheduler [2]. While the scheduler attempts to assign nearby nodes to jobs, no guarantees of locality are provided. As a result, task-mapping algorithms for Cray systems need to accommodate non-block, non-contiguous allocations. Our proposed methods can be applied to non-contiguous allocations as well as block-based allocations.

Ideally, closer proximity of router coordinates results in lower communication costs. However, congestion caused by communication patterns within an application and by other applications on the system can affect application behavior. In [21], information about input and output message stalls (a measure of network congestion [37]) was obtained from Cray Gemini tile counters and used to validate a computed congestion metric (Eqn. 7). Using this stalls data for graph-based mapping was explored in [14].

## 3 MAPPING METRICS

We define several metrics to evaluate our mappings. We assume static routing of messages. Also, we assume that each message is transferred over a single path (i.e., messages are not split and sent through multiple paths).

Let $G_t(V_t, E_t)$ be the *task communication graph*, where $V_t$ is the set of tasks, and $E_t$ is the set of edges that represent communication between tasks. If $t_1, t_2 \in V_t$, edge $(t_1, t_2) \in E_t$ if and only if tasks $t_1$ and $t_2$ communicate. The volume (weight) of the communication message is denoted with $w(t_1, t_2)$. In the same way, let $G_n(V_n, E_n)$ be the *network topology graph*. $V_n$ is the set of nodes in a machine, and $E_n$ is the set of edges that represent the physical communication links between nodes. If $n_1, n_2 \in V_n$, edge $e = (n_1, n_2) \in E_n$ if and only if nodes $n_1$ and $n_2$ have a connecting link between them. Each link $e$ is associated with a bandwidth that is denoted with $bw(e)$. Let $\mathcal{M}$ be a function for the assignment of tasks to nodes. That is, $n_1 = \mathcal{M}(t_1)$, if $t_1 \in V_t$ is assigned to node $n_1 \in V_n$.

Using these two graphs, we define the $Hops(t_1, t_2)$ of an edge $(t_1, t_2) \in E_t$ as the length of the shortest path between nodes $\mathcal{M}(t_1)$ and $\mathcal{M}(t_2)$ in the network topology graph $G_n$ (i.e., the number of links or "hops" a message from $t_1$ to $t_2$ travels in $G_n$). Then the total hops for any task assignment $\mathcal{M}$ is the sum of the hops for all the communicating tasks:

$$Hops(\mathcal{M}) = \sum_{(t_1, t_2) \in E_t} Hops(t_1, t_2). \tag{1}$$

When all messages have the same size (i.e., uniform $w(t_1, t_2)$ for all edges $(t_1, t_2)$), we examine the average number of edges traversed by each message:

$$AverageHops(\mathcal{M}) = Hops(\mathcal{M})/|E_t|. \tag{2}$$

With non-uniform $w(t_1, t_2)$, we measure weighted hops:

$$WeightedHops(\mathcal{M}) = \sum_{(t_1, t_2) \in E_t} w(t_1, t_2) Hops(t_1, t_2). \tag{3}$$

The total data on an edge $e \in E_n$ is defined as

$$Data(e) = \sum_{(t_1, t_2) \in E_t} w(t_1, t_2) \times InPath(e, \mathcal{M}(t_1), \mathcal{M}(t_2)), \tag{4}$$

where $InPath$ evaluates to 1 if and only if $e$ is in the shortest path between nodes $\mathcal{M}(t_1)$ and $\mathcal{M}(t_2)$, and 0 otherwise.

The maximum amount of data going through any link, then, is a measure of the congestion in the network:

$$Data(\mathcal{M}) = \max_{e \in E_n} Data(e) \tag{5}$$

$Data(\mathcal{M})$ ignores the bandwidth $bw(e)$ of network links $e$. We define serialization latency of an edge $e$ and an assignment $\mathcal{M}$ to account for links that are not uniform:

$$Latency(e) = \frac{Data(e)}{bw(e)}; \tag{6}$$

$$Latency(\mathcal{M}) = \max_{e \in E_n} Latency(e). \tag{7}$$

$Latency(e)$ is the time to transfer the given data through a given link $e$, while $Latency(\mathcal{M})$ is the time for the bottleneck link. Contention is an emperical measure of the number of stalls due to heavy traffic on the links. The contention of link $e$ is proportional to $Latency(e)$; the maximum contention in a network is proportional to $Latency(\mathcal{M})$.

## 4 GEOMETRIC TASK MAPPING

Our proposed topology-aware mapping algorithm uses the router coordinates to represent the network topology of the machine. The cost of communication between pairs of cores is approximated by the length of the shortest path between their routers' coordinates. Thus, the machine topology is described only by the cores' coordinates, rather than a topology graph in which bandwidth information between every pair of cores must be specified. Each of the application's MPI processes is also represented by a coordinate, corresponding to either the center of the process' application domain or the average coordinate of its application data. For example, in a structured grid-based finite difference application, the center of an MPI process' subgrid can be used as its coordinate. Our algorithm uses a geometric partitioning algorithm to consistently reorder both the MPI processes and the allocated cores; this reordering is used to construct the mapping. In this section, we provide details of our mapping algorithm. We use the term "machine coordinates" to refer to the router coordinates associated with each core, and "task coordinates" to refer to the centroid or averaged coordinates provided by the application's MPI processes.

### 4.1 Multi-Jagged (MJ) Geometric Partitioning

Our proposed task mapping algorithm uses a geometric partitioning algorithm, the Multi-dimensional Jagged algorithm (MJ) [20] of the Zoltan2 Toolkit [13], to partition task and machine coordinates. MJ partitions a set of coordinates into a desired number of parts ($P$) in a given number of steps called the *recursion depth* ($RD$). During each recursion, one-dimensional partitioning is done along a dimension; the partitioning dimension changes in each recursion. Therefore, MJ is a generalization of the Recursive Coordinate Bisection (RCB) algorithm [6] in which MJ has ability to do multisections instead of bisections. Although our implementation of MJ can partition into any number of parts $P$, we simplify our explanation here by assuming $P$ can be written as $P = \prod_{i=1}^{RD} P_i$. In the first level, MJ partitions the domain into $P_1$ parts using $P_1 - 1$ cuts in one direction. In the next level, each of the $P_1$ parts is partitioned separately into $P_2$ parts using cuts in an orthogonal direction. This recursion continues in each level. Figure 1 shows two 64-way partitions using MJ with $RD = 3$, $P = 4 \times 4 \times 4$ (left), and $RD = 6$, $P = 2 \times 2 \times 2 \times 2 \times 2 \times 2$ (right). When $RD = \lceil \log_2 P \rceil$, MJ is equivalent to RCB (as in Figure 1(b)).

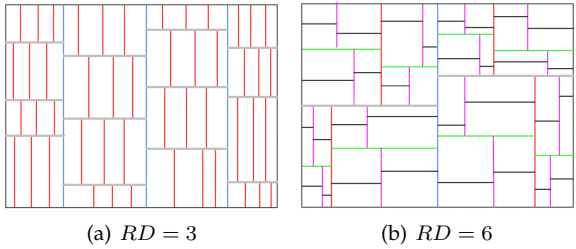

(a) $RD = 3$      (b) $RD = 6$

Fig. 1. Partitioning into 64 parts using MJ with different recursion depths. Cutlines in the same level of recursion share the same color.

MJ's complexity depends on $P$, $RD$, the number of points $n$, and the average number of iterations $it$ needed to compute cutline locations. During partitioning on level $i$, each point is compared to $\log_2 P_i$ cut lines (using binary search). Thus, MJ's complexity is $O(n \times it \times \sum_{i=1}^{RD} \log_2 P_i)$. When MJ is used as RCB, its complexity is $O(n \times it \times \log_2 P)$.

### 4.2 Using MJ for Task Mapping

Although MJ is proposed as a parallel (MPI+OpenMP) algorithm [20], we use it as a sequential algorithm in this context. The size of the partitioning problem is proportional to the number of processors. Since current supercomputers typically have $O(100K)$ processors, the partitioning algorithm would be communication bound if done in parallel; little or no speedup would be obtained by parallelizing this process. Instead, each processor calculates the mapping independently. A gather operation is performed at the beginning of task mapping to provide all machine and task coordinates to every processor. Then every processor performs the sequential mapping operation. We describe in Section 4.3 how we exploit parallelism to improve the quality of the mapping with minimal additional cost.

The mapping algorithm is defined as follows: Given $td$-dimensional coordinates of tasks ($tcoords$), and $pd$-dimensional coordinates of allocated cores ($pcoords$), along with the number of tasks ($tnum$) and cores ($pnum$), compute a mapping from tasks to cores ($\mathcal{M}$) and/or from cores to tasks ($\mathcal{M}^{-1}$). Algorithm 1 provides details.

---

**Algorithm 1** Task Mapping Algorithm using MJ

**Require:** $tcoords, tdim, tnum, pcoords, pdim, pnum, RD$

  NP ← MIN ($pnum, tnum$)
  TASKPARTITION ← MJ ($tcoords$, $tdim$, $tnum$, NP, $RD$)
  PROCPARTITION ← MJ ($pcoords$, $pdim$, $pnum$, NP, $RD$)
  $\mathcal{M}, \mathcal{M}^{-1}$ ← GETMAPPINGARRAYS ( $tnum$, $pnum$,
    TASKPARTITION , PROCPARTITION )

---

MJ's main purpose in Algorithm 1 is to order the tasks and processors in a way to exploit their hierarchies. Function MJ partitions the task and cores into NP parts, and assigns a part number to each task and core. Tasks and cores that share the same part number are then mapped to each other by GETMAPPINGARRAYS; the resulting mappings are stored in $\mathcal{M}$ and $\mathcal{M}^{-1}$.

There are three mapping cases, depending on $tnum$ and $pnum$:

**1)** $tnum = pnum$**:** A one-to-one mapping exists such that, for task $t$ assigned to core $p$, $t = \mathcal{M}^{-1}[p]$ and $p = \mathcal{M}[t]$.

**2)** $tnum > pnum$**:** When there are more tasks than cores, a core is assigned multiple tasks. Both cores and tasks are partitioned into $pnum$ parts, with multiple tasks in the each part. The mapping results will be $t \in \mathcal{M}^{-1}[p]$ and $p = \mathcal{M}[t]$. This case can be considered a one-phase mapping and partitioning algorithm that can perform simultanous partitioning and mapping.

**3)** $tnum < pnum$**:** When there are more cores than tasks, the algorithm does not split a task among multiple cores. Instead, it chooses a subset of $tnum$ cores. Then, mapping is performed within this subset as if $pnum = tnum$. Some cores will be idle, as they are not assigned any tasks. Our implementation uses a modified K-means clustering algorithm [26] to choose the closest subset of cores within the allocation. In this paper, this special case is not considered as it does not arise frequently in applications.

The complexity of Algorithm 1 is dominated by the calls to MJ, since $getMappingArrays$ runs in linear time with respect to $tnum$ and $pnum$. Thus, when MJ is used as RCB and $tnum = pnum$, the overall complexity of the mapping algorithm is $O(tnum \times it \times \log_2(tnum))$.

### 4.3 Improving the quality of the mapping

The ability of our mapping strategy to reduce communication costs depends on the results of the MJ partitioner. In this section, we describe several ways that we can improve the quality of the mapping by modifying the input to MJ. These improvements are computed with very little extra expense, as they are computed in parallel across sets of processors.

**Shifting the machine coordinates:** The first improvement involves considering the torus interconnection present in many supercomputer networks. Torus networks provide wrap-around communication links in each network dimension that are not reflected in the machine coordinates. Thus, since MJ is not aware of connectivity information, MJ considers nodes at edges of the network coordinates to be far apart, even though there is a one-hop path between them. We transform the coordinates to account for wrap-around in each dimension. Our shifting strategy applies a one-dimensional operation to each dimension independently. First, we find the shift position – the largest gap in the node coordinates. Then, assuming the largest gap is greater than one, we transform the machine coordinates on one side of the shift position by adding to them the maximum machine coordinate in that dimension. Refer to [21] for details.

**Rotating the machine and task coordinates:** The quality of the mapping also depends on the order of the dimensions to which the partitioning is applied (e.g., first partition in $x$, then $y$, then $z$). It is difficult to predict which dimension order will provide the best mapping. Since there are $pnum$ processes, we instead calculate different mappings with different rotations in each process. Then, given the communication pattern of the tasks, $WeightedHops$ (Eqn. 3) for each mapping is computed; the mapping with the smallest value is chosen. This comparison requires one extra Allreduce and broadcast operation. With $td$-dimensional tasks and $pd$-dimensional processors, there are $rp = (td)! \times (pd)!$ different rotations. For a 3D torus with 3D task coordinates, $rp = 3! \times 3! = 36$. We group processes into sets of size 36, in which each process calculates a mapping using a different rotation, along with the quality of that mapping. Then within each group, the best quality mapping is determined, and is broadcast to the group. Refer to [21] for details.

**Partitioning along the longest dimension:** In geometric methods, partitioning perpendicular to the longest dimension given by the data's coordinates is commonly used to minimize the surface area of the cutting plane, and, as a result, the communication volume between the partitions (e.g., RCB [6], Recursive Inertial Bisection [39]). Our previous work [21] used a fixed ordering of the partitioning dimensions (strictly alternating in each recursion level) rather than considering the longest dimension. It also assumed dimension sizes $pd$ and $td$ were equal; when $pd \neq td$, the smaller dimension was chosen. These decisons were made to achieve consistent ordering of the resulting parts. However, in order to find the task and node hierarchies, these decisions needed to be revisited.

In this work, we incorporate longest dimension partitioning into our task mapping algorithm to increase locality and connectivity between interdependent tasks and nodes. For example, Figure 2 shows the partitions of the node coordinates, without (left) and with (right) longest dimension partitioning. In Figure 2(left), strictly alternating partitioning directions are used, starting with a vertical cut (gray), then a horizontal cut (blue), and another vertical cut (red). The resulting parts $p0$-$p3$ are assigned to nodes $n0$-$n3$, which are connected with three links in a mesh network. With longest direction partitioning, parts $p0$-$p3$ are assigned to to nodes $n0$, $n8$, $n1$ and $n9$, which are connected by four links. (Equal benefits are seen for the other parts in this example.) Thus, the longest-dimension partition provides higher bandwidth for the highly communicating tasks.

**Adaptation of space filling orderings for part numbers:** MJ assigns part numbers using Z-order (Z) space filling curves [36]. That is, it recursively assigns lower part numbers to parts with coordinates less than the cut coordinate. Figure 3 shows several space filling curve orderings in the literature, applied into 64 parts. Hilbert's order (H) is achieved by partitioning into quadrants recursively; a rotation of the coordinates (e.g., swapping $x$ and $y$ coordinates) or flip of the coordinates (i.e, multiplying each coordinate by $-1$) is applied to different quadrants. Similarly, Gray order (G) is obtained by bisecting the dataset recursively, and flipping all coordinates of the higher part. In this work, we use a Flipped Z order (FZ), which is achieved by flipping one coordinate of each points in the higher part; it is similar to Gray and Z ordering, but the flipping is applied only to the coordinate for which the cut is obtained. Algorithm 2 shows the modified MJ that adopts these orderings. To simplify the presentation, the MJ algorithm shown performs recursive bisection, and its target number of parts is assumed to be a power of two; these restrictions do not apply in the actual implementation. In each recursion, the algorithm

**Algorithm 2** MULTI-JAGGED Algorithm (MJ). For simplicity, the algorithm shown runs using bisection and assumes the number of parts $np$ is a power of two.

---

**Require:** $coords_{dim,ncoord}, dim, ncoord, np$
  $\mu(c) \leftarrow 0$, for $0 \leq c < ncoord$
  //*sfc type is Z, Gray or FZ*
  MJ_Helper ($coords_{*,*}, dim, ncoord, np, sfc, \mu$)
  **return** $\mu$

---

*procedure* MJ_HELPER($coords_{*,*}, dim, nc, np, sfc, \mu$)
  **if** $np = 1$ **then**
      return
  **end if**
  //get longest dimension
  $d \leftarrow$ GETPARTDIM($coords_{*,*}, ncoord, dim$)
  //in dim $d$, bisect into parts $L$ and $R$
  $L, R \leftarrow$ BIN1DPART($coords_{d,*}, ncoord$)
  $np \leftarrow \frac{np}{2}$
  **for** $r \in R$ **do**
      **if** *sfc is Gray* **then**
          $coords_{*,r} \leftarrow -coords_{*,r}$
      **else if** *sfc is FZ* **then**
          $coords_{d,r} \leftarrow -coords_{d,r}$
      **end if**
      $\mu(r) \leftarrow \mu(r) + np$
  **end for**
  MJ_Helper($coords_{*,L}, dim, \|L\|, np, sfc, \mu$)
  MJ_Helper($coords_{*,R}, dim, \|R\|, np, sfc, \mu$)
*end procedure*

---

finds the dimension to partition and calls a 1D partitioning algorithm. Z ordering does not modify the coordinates; it always assigns lower part numbers to coordinates less than the cut position, and higher part numbers to coordinates greater than the cut. Gray ordering flips the coordinates in all dimensions for points in the higher half, while the proposed FZ ordering flips only the coordinates for the dimension in which the bisection was performed.

Our previous work [21] enforced consistent cut order for both task and node partitions, even for datasets in which the dimensions of task and node coordinates differed. When the task and node dimensions differed, the minimum dimension was used. Thus, all mappings using any ordering from Figure 3 were equivalant. In this work, we loosen the restriction of equal task and node dimensions to support high-dimensional networks like the 5D network in Blue-Gene/Q. In such cases, the choice of ordering affects the quality of the mapping because the cut order is different in the task and node partitions.

As we will show in the experiments, FZ ordering obtains superior performance in general than Z ordering. (Here, we briefly explain the the differences; more detailed analysis is in Appendix A.) FZ has advantages over Z when

- the ordering of the cut dimensions is not consistent between the task and node networks, or
- the task or node network has wrap-around links.

For example, we show 3D and 1D partitions of 64 points using FZ and Z orderings in Figures 4 and 5. We then map 2D task graphs (as given in Figure 3) to the networks in

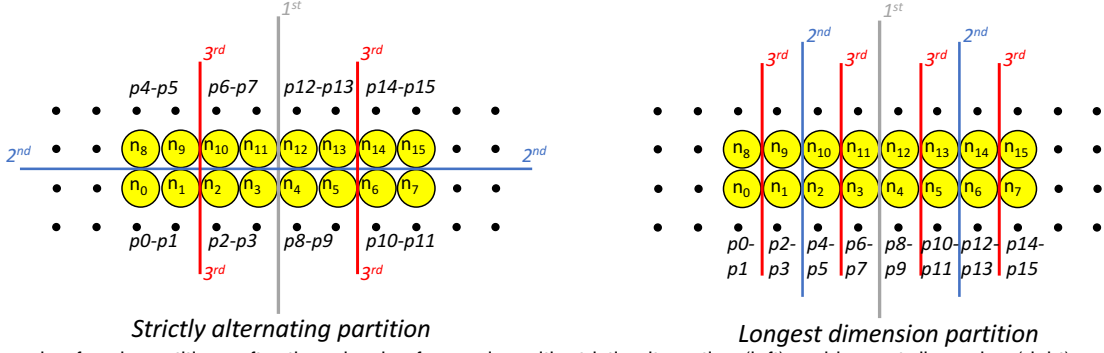

Fig. 2. An example of node partitions after three levels of recursion with strictly alternating (left) and longest dimension (right) partitioning. Gray, blue, and red lines show the first, second, and third dimension cut-lines, respectively.

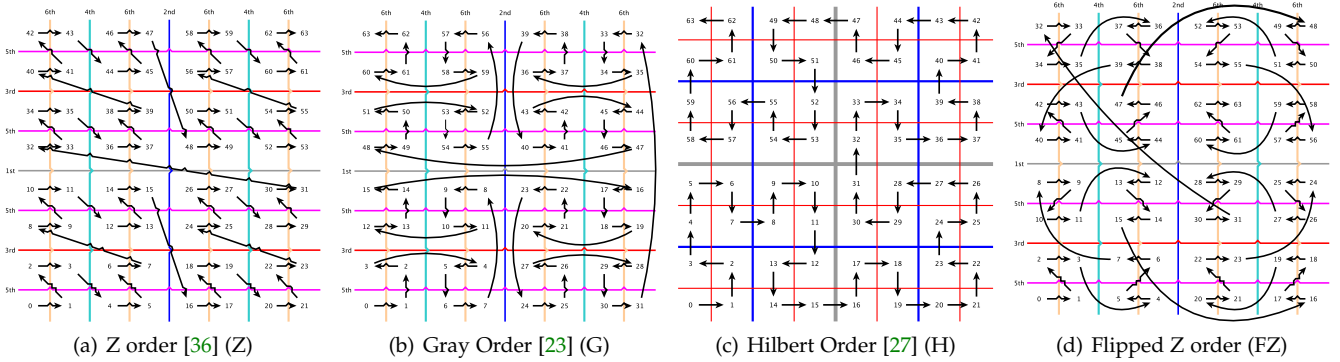

(a) Z order [36] (Z)  (b) Gray Order [23] (G)  (c) Hilbert Order [27] (H)  (d) Flipped Z order (FZ)

Fig. 3. Examples of different space-filling curve orderings of partitions. For (a), (b), and (d), the order of the cuts is indicated with colors; the first to last cuts are represented by gray, blue, red, cyan, orange and purple lines, respectively.

Figures 4 and 5. We assume that each task communicates with only its immediate neighbors, and each node has links with only its immediate neighbors along each dimension.

When Z ordering is used to partition 2D tasks (Figure 3(a)) and 3D nodes (Figure 4(a)):

- The first gray cut separates tasks $(10, 32)$, $(11, 33)$, $(14, 36)$, $(15, 37)$, ...; messages to corresponding pairs travel $(2, 2, 1)$ hops in the $(x, y, z)$ dimensions.
- The second blue cuts separate tasks $(47, 58)$, $(45, 56)$, $(39, 50)$, $(37, 48)$, ..., with $(1, 1, 2)$ hops.
- The third red cuts separate tasks $(34, 40)$, $(35, 41)$, $(38, 44)$, $(39, 45)$, ..., with $(2, 0, 1)$ hops.
- The fourth cyan cuts separate tasks $(43, 46)$, $(41, 44)$, $(35, 38)$, $(33, 36)$, ..., with $(1, 1, 0)$ hops.
- The fifth purple cuts separate tasks $(40, 42)$, $(41, 43)$, $(44, 46)$, $(45, 47)$, ..., with $(0, 0, 1)$ hops.
- The sixth orange cuts separate tasks $(42, 43)$, $(40, 41)$, $(34, 35)$, $(32, 33)$, ..., with $(1, 0, 0)$ hops.

Each task has four neighbors, and although the 3D node network provides more links per node, only the task neighbors that are separated along the last two cuts are placed in adjacent nodes. For example, node 35 is adjacent to nodes 33, 34, 42, 49, 7 and 39. For task 35, it communicates with nodes 33 and 34, while task 35's other neighboring tasks 38 and 41 are placed two and three hops away.

Again using 2D task and 3D node networks, Figures 3(d) and 4(b) show the partitions using FZ ordering. In this case,

- The first gray cut separates tasks $(8, 40)$, $(9, 41)$, $(13, 45)$, $(12, 44)$, ...; messages to corresponding

pairs travel either 1 (if there are wrap-around links) or 3 (if not) hops in the $y$ dimension.
- The second blue cuts separate tasks $(36, 52)$, $(38, 54)$, $(46, 62)$, $(44, 60)$, ..., with either 1 (with wrap-around) or 3 (without) hops along $z$.
- The third red cuts separate tasks $(34, 42)$, $(35, 43)$, $(39, 47)$, $(38, 46)$, ..., with either 1 (with wrap-around) or 3 (without) hops along $x$.
- The fourth cyan cuts separate tasks $(33, 37)$, $(35, 39)$, $(43, 47)$, $(41, 45)$, ..., with 1 hop along $y$.
- The fifth purple cuts separate tasks $(32, 34)$, $(33, 35)$, $(37, 39)$, $(36, 38)$, ..., with 1 hop along $z$.
- The sixth orange cuts separate tasks $(32, 33)$, $(34, 35)$, $(42, 43)$, $(40, 41)$, ..., with 1 hop along $x$.

Task pairs separated with the last three cuts are all placed one hop apart. For the other cuts, the number of hops increases from one hop to $pd - 1$ hops. The number of hops is reduced when there are torus wrap-around links; with wrap-around links, each message is sent using only a single hop. A node in the network, such as 39, is next to the nodes 7, 35, 37, 38, 47, 55, where each neighbor differs in a single bit as a Gray-code property. All of task 39's neighbors, 35, 37, 38 and 47, are placed in neighboring nodes.

In this example, we have different cut-dimension ordering for tasks and nodes, as their networks have different dimensions. We have shown that FZ ordering outperform Z in such cases. The same relative performance results when $3D$ tasks are mapped to a $2D$ node network. Also, in these examples, we alternated the cut dimension for each cut. With the use of longest dimension partitioning, the cut-dimension orders are more likely to be arbitrary even for datasets with matching dimensions. In such cases, FZ ordering is expected to have even greater advantage over Z.

In general, Z ordering achieves good performance when $td \neq pd$ and $td$ is a multiple of $pd$. For example, consider mapping the 2D tasks to a 1D processor network. 1D ordering with Z and FZ is shown in Figure 5. With Z, messages from task 44 to its neighbors travel 3, 2, 1, and 6 hops; with FZ, messages to task 44's neighbors travel 1, 3, 15 and 47 hops. However, even in this case, the addition of torus links improves FZ ordering. For example, with Z order, the number of hops to neighbors of task 37 is 1, 2, 11 and 22; with FZ, task 37's neighbors are 5, 13, 1 and 1 hops away.

Another example of the structured case occurs when $pd \neq td$ and $pd$ is a multiple of $td$, such as when the previous example is reversed so that 1D tasks are mapped to 2D nodes. For example, with Z-order, the tasks that are separated along the cuts corresponding to the third level of recursion are $(7, 8)$, $(23, 24)$, $(39, 40)$, and $(55, 56)$. Messages corresponding to these pairs travel 3 hops along $x$ and 1 along $y$. If FZ ordering is used, the same cuts separate tasks $(4, 12)$, $(28, 20)$, $(52, 60)$, and $(44, 36)$; their messages travel 3 hops along $y$. Although FZ already reduces the hops relative to Z, we can reduce the hops further when $pd$ is a multiple of $td$. For these cases, we propose MFZ, a slightly modified FZ. In MFZ, we number only one of the coordinate sets (either tasks or nodes) using Z ordering. When we partition the other set, we flip the coordinates in the lower half, rather than the higher half. For example, application of MFZ to a 1D dataset is shown in Figure 5. With mapping of 1D tasks with MFZ ordering to 2D nodes using FZ as in Figure 3(d), the third-level cuts divide tasks $(27, 28)$, $(3, 11)$, $(43, 35)$, and $(51, 59)$. Each of these tasks are separated by only one hop along $y$ in Figure 3(d). Therefore, MFZ is a modification of FZ that differs from FZ only when $pd$ is a multiple of $td$ to reduce the number of hops further. Appendix A provides detailed analysis of Z and FZ.

## 5 EXPERIMENTS

We evaluate the proposed geometric mapping methods. In Section 5.1, we compare the quality of different orderings in terms of $AverageHops$ (Eqn. 2) with several different task- and node-coordinate dimensions. The proposed task mapping method is evaluated on a BlueGene/Q network (Mira) in Section 5.2, and on a Cray Gemini network (Titan) in Section 5.3. The Gemini results are applicable to other Gemini networks such as BlueWaters; those on BlueGene/Q are also applicable to Vulcan and Sequoia. See [21] for experiments of our earlier methods on two other Cray Gemini interconnection networks (Hopper and Cielo).

### 5.1 Effect of Ordering on Mapping Quality

We study the effect of the SFC orderings on the quality of geometric mapping. We generate $td$-dimensional mesh- and torus-connected tasks, where tasks communicate with their immediate neighbors. These tasks are then one-to-one mapped to $pd$-dimensional block-allocated nodes. Table 1 gives the calculated $AverageHops$ for these mappings using the Hilbert (H), Z-order (Z), and Flipped Z-order (FZ) curves. When $pd$ is a multiple of $td$, Modified Flipped Z-order (MFZ) is applied as well. Results with Gray-order (G) (not shown) were similar to Z-order results.

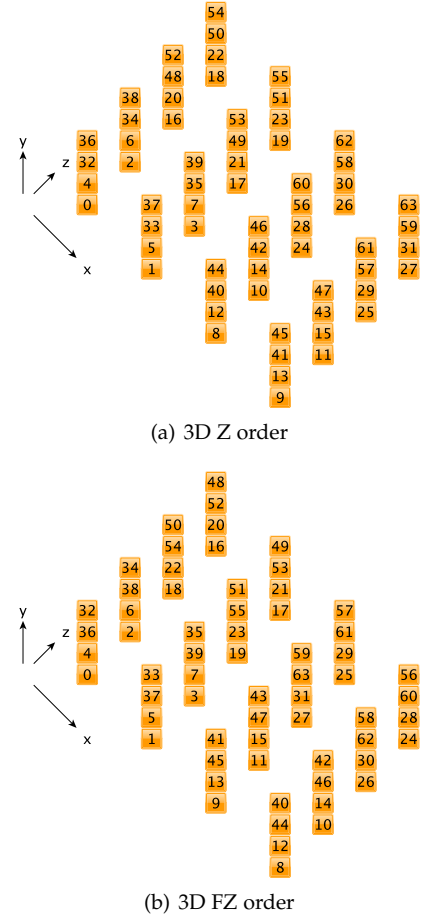

(a) 3D Z order

(b) 3D FZ order

Fig. 4. Examples of Z and FZ space-filling curves for 3D datasets.

In general, Hilbert and Flipped Z-order are better than Z-order and Gray. When $td$ or $pd$ is one, Hilbert obtains the lowest $AverageHops$. Because Hilbert is a continous ordering, it does not have any jumps when mapping 1D to multi-dimensional orderings. As explained in Section 4.3, when $td$ is a multiple of $pd$, Z-order obtains the lowest $AverageHops$. For the rest of the experiments, Flipped Z-order is significantly better than both Hilbert and Z-order. $AverageHops$ are reduced further with Modified Flipped Z-order (MFZ) on datasets with $pd$ (mod $td$) = 0. On average, MFZ obtains 27% and 38% lower $AverageHops$ than Z-order and Hilbert orderings; it also improves upon Flipped Z-order by 7%. Because of the cyclic properties of the Gray Ordering, Flipped Z-order captures locality in torus datasets; therefore, its reductions are even greater when nodes and tasks have torus connectivity. The performance of Z-order is similar on mesh and torus-based processors because it does not exploit the cyclic properties of the networks. As Flipped Z-order is the best in most cases, we use this ordering for the rest of the experiments.

### 5.2 Task Mapping for BlueGene/Q

We evaluate the mapping methods using an atmospheric modeling application, HOMME (High-Order Method Mod-

| Z | 0 | 1 | 2 | 3 | 4 | 5 | 6 | 7 | 8 | 9 | 10 | 11 | 12 | 13 | 14 | 15 | 16 | 17 | 18 | 19 | 20 | 21 | 22 | 23 | 24 | 25 | 26 | 27 | 28 | 29 | 30 | 31 | 32 | 33 | 34 | 35 | 36 | 37 | 38 | 39 | 40 | 41 | 42 | 43 | 44 | 45 | 46 | 47 | 48 | 49 | 50 | 51 | 52 | 53 | 54 | 55 | 56 | 57 | 58 | 59 | 60 | 61 | 62 | 63 |
|---|---|---|---|---|---|---|---|---|---|---|---|---|---|---|---|---|---|---|---|---|---|---|---|---|---|---|---|---|---|---|---|---|---|---|---|---|---|---|---|---|---|---|---|---|---|---|---|---|---|---|---|---|---|---|---|---|---|---|---|---|---|---|---|
| **FZ** | 0 | 1 | 3 | 2 | 6 | 7 | 5 | 4 | 12 | 13 | 15 | 14 | 10 | 11 | 9 | 8 | 24 | 25 | 27 | 26 | 30 | 31 | 29 | 28 | 20 | 21 | 23 | 22 | 18 | 19 | 17 | 16 | 48 | 49 | 51 | 50 | 54 | 55 | 53 | 52 | 60 | 61 | 63 | 62 | 58 | 59 | 57 | 56 | 40 | 41 | 43 | 42 | 46 | 47 | 45 | 44 | 36 | 37 | 39 | 38 | 34 | 35 | 33 | 32 |
| **MFZ** | 3 | 1 | 0 | 2 | 9 | 8 | 29 | 25 | 26 | 27 | 19 | 16 | 17 | 22 | 20 | 23 | 7 | 6 | 4 | 5 | 11 | 10 | 14 | 15 | 13 | 12 | 28 | 30 | 24 | 31 | 21 | 18 | 35 | 39 | 55 | 51 | 48 | 49 | 59 | 55 | 57 | 56 | 40 | 41 | 43 | 42 | 52 | 53 | 49 | 48 | 50 | 51 | 58 | 57 | 54 | 45 | 44 | 47 | 46 | 60 | 62 | 61 | 63 |

Fig. 5. Examples of Z, FZ and MFZ space-filling curves for 1D datasets.

TABLE 1

$AverageHops$ (Eqn. 2) resulting from geometric mapping with different SFC orderings: Hilbert (H), Z-order (Z), Flipped Z-order (FZ), and Modified Flipped Z-order (MFZ). Each row represents a mapping with $td$-dimensional tasks onto $pd$-dimensional nodes. The left-most column gives the number of tasks and nodes used. Along each dimension, the number of tasks (similarly, nodes) is equal. Gray order obtains very similar quality to Z-order; therefore, we do not show results for Gray. The `A to B` headings indicate that A-connected tasks are mapped onto a B-connected network. For example, the results for mapping mesh-connected tasks onto a torus-connected network have the heading "Mesh To Torus." MFZ's modification is used only when $pd$ is a multiple of $td$; otherwise, MFZ is identical to FZ. The best method in each instance is highlighted in red.

| # task | pd | td | Mesh to Mesh | | | | Mesh to Torus | | | | Torus to Torus | | | |
|---|---|---|---|---|---|---|---|---|---|---|---|---|---|---|
| | | | H | Z | FZ | MFZ | H | Z | FZ | MFZ | H | Z | FZ | MFZ |
| 262,144 | 1 | 2 | 311.05 | 256.50 | 384.00 | | 246.92 | 256.50 | 351.94 | | 411.01 | 426.67 | 447.25 | |
| 32,768 | | 3 | 380.49 | 352.33 | 410.67 | | 292.40 | 352.33 | 322.58 | | 518.62 | 633.90 | 525.83 | |
| 1,048,576 | | 4 | 8755.69 | 8456.25 | 9060.00 | | 6641.63 | 8456.25 | 6945.94 | | 12324.09 | 15837.86 | 12360.88 | |
| 32,768 | | 5 | 951.63 | 936.20 | 967.20 | | 717.57 | 936.20 | 733.14 | | 1229.50 | 1611.95 | 1230.30 | |
| 262,144 | | 6 | 6291.69 | 6241.50 | 6342.00 | | 4731.31 | 6241.50 | 4781.62 | | 8193.25 | 10835.96 | 8194.58 | |
| 65,536 | | 8 | 2735.92 | 2730.63 | 2741.25 | | 2053.25 | 2730.63 | 2058.58 | | 3071.94 | 4087.94 | 3071.94 | |
| 262,144 | 2 | 1 | 1.00 | 2.00 | 1.99 | 1.20 | 1.00 | 1.99 | 1.99 | 1.20 | 1.00 | 1.99 | 1.99 | 1.20 |
| 262,144 | | 3 | 11.55 | 13.45 | 10.67 | | 10.79 | 13.45 | 9.31 | | 14.03 | 17.81 | 11.17 | |
| 1,048,576 | | 4 | 24.63 | 16.50 | 24.00 | | 21.15 | 16.50 | 21.94 | | 32.93 | 26.66 | 27.25 | |
| 1,048,576 | | 5 | 40.11 | 39.92 | 34.56 | | 34.38 | 39.92 | 27.73 | | 53.28 | 62.20 | 40.40 | |
| 262,144 | | 6 | 31.22 | 24.33 | 28.00 | | 26.14 | 24.33 | 21.90 | | 41.43 | 39.58 | 32.50 | |
| 65,536 | | 8 | 25.73 | 21.25 | 22.50 | | 21.28 | 21.25 | 17.17 | | 30.59 | 30.88 | 23.88 | |
| 32,768 | 3 | 1 | 1.00 | 2.00 | 1.33 | 1.04 | 1.00 | 1.99 | 1.32 | 1.04 | 1.00 | 1.99 | 1.32 | 1.04 |
| 262,144 | | 2 | 2.56 | 3.30 | 1.97 | | 2.50 | 3.28 | 1.88 | | 2.55 | 3.40 | 1.89 | |
| 4,096 | | 3 | 3.46 | 3.54 | 2.57 | | 3.18 | 3.54 | 2.14 | | 3.80 | 4.50 | 2.38 | |
| 32,768 | | 5 | 5.33 | 5.11 | 3.89 | | 4.79 | 5.11 | 3.20 | | 6.10 | 6.80 | 3.80 | |
| 262,144 | | 6 | 7.15 | 4.50 | 6.00 | | 6.23 | 4.50 | 5.43 | | 8.97 | 6.63 | 6.25 | |
| 262,144 | | 9 | 9.89 | 7.00 | 7.78 | | 8.41 | 7.00 | 6.00 | | 11.67 | 9.83 | 7.83 | |
| 1,048,576 | 4 | 1 | 1.00 | 2.00 | 1.14 | 1.01 | 1.00 | 2.00 | 1.14 | 1.01 | 1.00 | 2.00 | 1.14 | 1.01 |
| 1,048,576 | | 2 | 1.80 | 1.94 | 1.91 | 1.17 | 1.80 | 1.91 | 1.82 | 1.17 | 1.82 | 1.91 | 1.82 | 1.18 |
| 4,096 | | 3 | 2.38 | 2.58 | 1.60 | | 2.21 | 2.58 | 1.38 | | 2.37 | 3.00 | 1.42 | |
| 1,048,576 | | 5 | 4.91 | 4.75 | 3.20 | | 4.61 | 4.75 | 2.77 | | 5.47 | 6.00 | 3.10 | |
| 4,096 | | 6 | 2.83 | 2.44 | 2.00 | | 2.48 | 2.44 | 1.56 | | 2.89 | 3.00 | 1.67 | |
| 65,536 | | 8 | 3.79 | 2.50 | 3.00 | | 3.24 | 2.50 | 2.67 | | 4.25 | 3.25 | 2.75 | |
| 32,768 | 5 | 1 | 1.00 | 2.00 | 1.07 | 1.00 | 1.00 | 1.99 | 1.06 | 1.00 | 1.00 | 1.99 | 1.06 | 1.00 |
| 1,048,576 | | 2 | 1.96 | 2.43 | 1.27 | | 1.94 | 2.42 | 1.24 | | 1.95 | 2.44 | 1.24 | |
| 32,768 | | 3 | 2.38 | 2.55 | 1.46 | | 2.27 | 2.55 | 1.31 | | 2.37 | 2.83 | 1.33 | |
| 1,048,576 | | 5 | 3.18 | 3.27 | 1.94 | | 3.03 | 3.27 | 1.74 | | 3.24 | 3.75 | 1.81 | |
| 1,048,576 | | 10 | 3.93 | 2.50 | 3.00 | | 3.36 | 2.50 | 2.67 | | 4.38 | 3.25 | 2.75 | |
| 262,144 | 6 | 1 | 1.00 | 2.00 | 1.03 | 1.00 | 1.00 | 2.00 | 1.03 | 1.00 | 1.00 | 2.00 | 1.03 | 1.00 |
| 262,144 | | 2 | 1.67 | 1.96 | 1.30 | 1.03 | 1.65 | 1.91 | 1.22 | 1.03 | 1.67 | 1.91 | 1.22 | 1.03 |
| 262,144 | | 3 | 1.91 | 1.78 | 1.67 | 1.10 | 1.84 | 1.68 | 1.38 | 1.10 | 1.91 | 1.69 | 1.38 | 1.13 |
| 4,096 | | 4 | 1.97 | 1.93 | 1.29 | | 1.77 | 1.93 | 1.00 | | 1.89 | 2.25 | 1.00 | |
| 262,144 | | 9 | 3.05 | 2.44 | 2.00 | | 2.67 | 2.44 | 1.56 | | 3.12 | 3.00 | 1.67 | |
| 65,536 | 8 | 1 | 1.00 | 2.00 | 1.01 | 1.00 | 1.00 | 1.99 | 1.00 | 1.00 | 1.00 | 1.99 | 1.00 | 1.00 |
| 65,536 | | 2 | 1.60 | 1.95 | 1.12 | 1.00 | 1.57 | 1.87 | 1.00 | 1.00 | 1.59 | 1.88 | 1.00 | 1.00 |
| 65,536 | | 4 | 1.74 | 1.60 | 1.40 | 1.00 | 1.60 | 1.47 | 1.00 | 1.00 | 1.73 | 1.50 | 1.00 | 1.00 |
| 262,144 | 9 | 1 | 1.00 | 2.00 | 1.00 | 1.00 | 1.00 | 2.00 | 1.00 | 1.00 | 1.00 | 2.00 | 1.00 | 1.00 |
| 262,144 | | 2 | 1.68 | 2.06 | 1.05 | | 1.64 | 2.06 | 1.00 | | 1.64 | 2.09 | 1.00 | |
| 262,144 | | 3 | 1.78 | 1.86 | 1.22 | 1.00 | 1.70 | 1.73 | 1.00 | | 1.74 | 1.75 | 1.00 | 1.00 |
| 262,144 | | 6 | 2.14 | 1.93 | 1.29 | | 1.88 | 1.93 | 1.00 | | 2.00 | 2.25 | 1.00 | |
| 1,048,576 | 10 | 1 | 1.00 | 2.00 | 1.00 | 1.00 | 1.00 | 2.00 | 1.00 | 1.00 | 1.00 | 2.00 | 1.00 | 1.00 |
| 1,048,576 | | 2 | 1.61 | 1.99 | 1.06 | 1.00 | 1.59 | 1.93 | 1.00 | | 1.59 | 1.94 | 1.00 | 1.00 |
| 1,048,576 | | 4 | 2.08 | 2.08 | 1.16 | | 1.92 | 2.08 | 1.00 | | 2.00 | 2.25 | 1.00 | |
| 1,048,576 | | 8 | 1.76 | 1.64 | 1.40 | 1.00 | 1.61 | 1.47 | 1.00 | | 1.74 | 1.50 | 1.00 | |
| GEOMEAN | | | 6.69 | 7.24 | 5.64 | 5.26 | 6.07 | 7.17 | 4.88 | 4.69 | 7.24 | 8.68 | 5.50 | 5.28 |
| Normalized w.r.t. Best | | | 1.27 | 1.38 | 1.07 | 1.00 | 1.30 | 1.53 | 1.04 | 1 | 1.37 | 1.64 | 1.04 | 1 |

eling Environment) [18], on the IBM BlueGene/Q computer Mira at Argonne National Laboratory. HOMME is part of the Energy Exascale Earth System Model (E3SM). It use an unstructured quadrilateral finite element mesh on a sphere, such as a cubed-sphere mesh for quasi-uniform resolution. Each surface element is projected into a column of hexahedral elements in the atmosphere. It employs spectral element and discontinous Galerkin methods to solve the shallow water or dry/moist primitive equations.

Figure 6 shows a simplified 2D slice of the assignment of mesh elements to processors in HOMME. A task (shown with a single color) is a single vertical column of the elements in the atmosphere. HOMME is designed to scale to one task per core. Thus, we run strong scaling experiments with various configurations. Tasks have 3D coordinates; the BlueGene/Q network has a 5D torus layout. The mapping methods that we compare are explained below:

• **SFC:** HOMME's default partitioning and mapping use Hilbert space filling curves. HOMME projects the sphere representing the earth onto a cube to obtain six faces, and partitions the tasks on these faces using Hilbert curves. The mapping is the output part number from the SFC; that is, an MPI rank is given the part with the same number. The rank ordering in BG/Q can be changed with built-in methods. ABCDET is the default mapping method; it places MPI ranks first along T (hardware threads within a node), then along E, then D, and so on. Other built-in mappings can be generated with different permutations such as TABCDE, TEABCD, etc. In our experiments, the HOMME SFC mappings with rank ordering ABCDET obtained the best results; thus, we report results using only ABCDET.

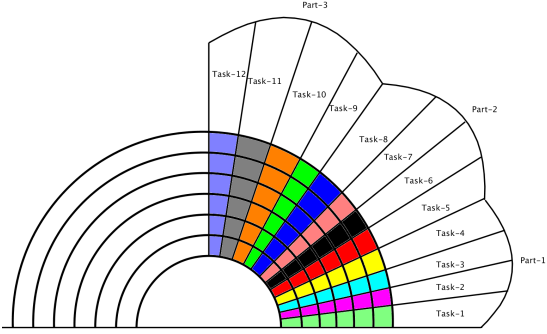

Fig. 6. A simple distribution of tasks to parts in a 2D slice of HOMME. The inner circle is the earth surface, and there are multiple elements in a vertical column in the layer of the atmosphere. Tasks correspond to vertical columns of elements.

● **SFC+Z2:** HOMME's SFC method is used to partition the mesh elements as in SFC. Then our proposed mapping method with FZ ordering (as implemented in Zoltan2) is used to map the parts to nodes.

● **Z2:** Our proposed mapping method with FZ ordering is used to both partition and map the mesh elements to processors within a single step.

In addition to the above variants, we investigated several application- and architecture-specific optimizations.

● **Architecture Specific Optimizations:** BlueGene/Q has a 5D torus network. A complete torus with different dimension lengths is allocated for each allocation. For example, allocations with 512 and 2048 nodes contain complete toruses, usually with dimensions $4 \times 4 \times 4 \times 4 \times 2$ and $4 \times 4 \times 4 \times 16 \times 2$, respectively. Allocations along the $E$ dimension can have length at most two. Therefore, there are two links that connect the immediate $E$ neighbors. Moreover, the network driver has special optimizations for the messages routed along $E$. Therefore, it is usually best to place heavily communicating tasks within a node, and then within nodes that are neighbors along $E$. We achieve this by ignoring $E$ during the partitioning of the processors. We denote this optimization in the figures below with "+$E$".

● **Application Specific Optimizations:** The application coordinates in HOMME are 3D coordinates on a sphere, as shown in Figure 7(a). HOMME's SFC implementation projects these coordinates to a cube 7(b) and partitions these cube coordinates. In our preliminary experiments, we saw that Z2 outperformed SFC in terms of mapping quality on smaller numbers of parts. However, as the number of parts increased, SFC obtained better quality than Z2. Z2 started by slicing the sphere; then partitioning the sphere slices became difficult in further steps when the number of parts was high. Partitioning the cube coordinates made achieving higher quality easier in further partitioning. Therefore, we transform the application coordinates in the pre-processing step as in Figure 7(b). In addition, in order to make use of the torus' wrap-around links, we transform the cube coordinates into 2D face coordinates, preserving the locality as much as possible as shown in Figures 7(c) and 7(d). With the 2D transformation of the task coordinates, the furthest mesh elements along the $x$ dimension are connected, which helps Z2 to place these tasks on nearby processors.

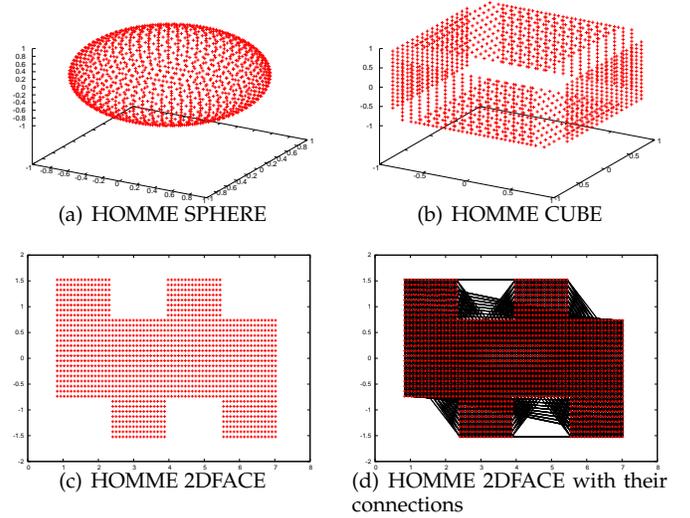

Fig. 7. The original sphere coordinates, transformed cube and 2D face coordinates of the mesh elements in HOMME.

HOMME is usually run with a hybrid setting in Mira; four MPI ranks are executed within a node, and each rank is occupied with 16 threads (with hyperthreads). In order to study strong scaling, we first run HOMME with MPI only (16 ranks per node); then we run the mapping in hybrid mode as in the real-use case. HOMME is usually run to simulate time frames as big as years; however, we simulate only a single day, since from the mapping perspective, the communication pattern does not change. We use $98,304$ tasks (six faces of the cube-sphere with $128 \times 128$ elements per face) and various numbers of nodes to study strong scaling. In the experiments, we use the lightweight BGQNCL library [9] to monitor the network link occupancies.

Table 2 gives the communication times of MPI-only HOMME with different mappings (SFC, SFC+Z2, and Z2) and coordinate transformations (Sphere, Cube and 2DFace from Figure 7). Times are normalized with respect to the communication time with HOMME's SFC mapping on $8K$ processors. With SFC, HOMME's communication time does not strong scale; there is a $36\%$ and $56\%$ reduction in the communication time on $16K$ and $32K$ processors. On $8K$ processors, neither SFC+Z2 nor Z2 variants (using Sphere, Cube and 2DFace transformations, with or without "+$E$" optimization) are able to reduce the communication time further. Although SFC+Z2 reduces $WeightedHops$ by 1-$25\%$, it increases $Data$ (Eqn. 5) by 14-31% on $8K$ processors (not shown). And even though Z2 reduces both average Data (Eqn. 4) per link and $WeightedHops$, it increases the number of messages exchanged in the system. We believe that these are the reasons for not seeing any reductions on $8K$ processors. On the other hand, as the number of processors increases, our mapping methods reduce the com-munication time significantly. For example, SFC+Z2 and Z2 reduce the communication time by 17% (16%) and 20% (27%) on $16K$ ($32K$) processors, respectively. Z2 usually performs best with 2DFace coordinates, but using 2DFace coordinates for SFC+Z2 hurts performance, different coordinate transformations are used during the partitioning and mapping steps. The architecture-specific "$+E$" optimization usually reduces communication time by another $4 - 5\%$ (and at most 22%). Overall, the communication time of HOMME with Z2 is reduced by 49% and 67% on $16K$ and $32K$ processors with respect to $8K$ processors. Thus, strong scaling for communication time is improved.

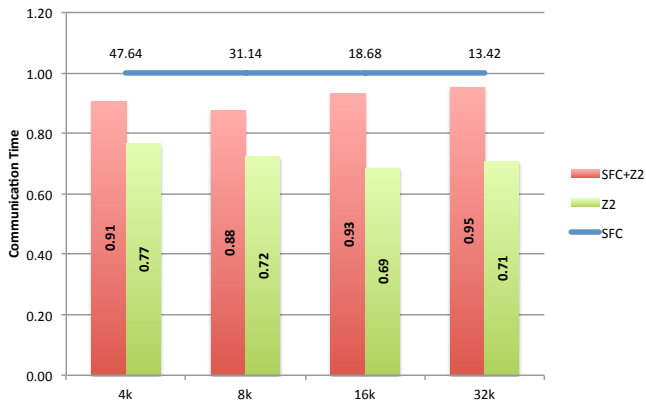

**Fig. 8.** Hybrid HOMME communication time normalized with respect to the time using HOMME's SFC mapping with 4K ranks. Actual communication time for HOMME with SFC are given above the blue line.

Figure 8 shows communication time in hybrid HOMME normalized with respect to the time using HOMME using its default SFC mapping on $4K$ ranks. In this experiment, we examine strong scaling using 1024 to 8192 nodes with four ranks each. At low node counts, each rank has 12 threads; however, above 4096 nodes, we reduce the number of threads because we run out of tasks (98304 tasks in total). We run only the best variants of our mappings for SFC+Z2 (Cube+E) and Z2 (2DFace+E). SFC+Z2 and Z2 reduce communication time up to 12% and 31%, respectively. Figure 9 gives the maximum and average amounts of data across the A, B, C, D, and E links, as measured with BGQNCL. Since all links have uniform bandwidth, link contention is proportional to these values. As seen in Figure 9(a), our mapping methods reduce $Data$ (Eqn. 5). Figure 9(b) shows that they also improve utilization of the links. HOMME with SFC overutilizes the D and E links, while it underutilizes A, B and C. Highly communicating tasks are given consecutive part numbers with SFC; therefore, with the ABCDE rank ordering, most of the communication occurs along D and E. The SFC+Z2 and Z2 mapping methods distribute the communication along the other dimensions, providing better link utilization as well as less contention in the links.

### 5.3 Task Mapping for Gemini Interconnection Networks

In this section, we evaluate the mapping methods on HOMME and a proxy application, MiniGhost [5], on the Cray XK7 (Titan) at Oak Ridge National Laboratory.

#### 5.3.1 HOMME Task Mapping on Titan

We evaluate our task mapping with a strong scaling experiment using HOMME on Titan. The size of the dataset is $86,400$ surface elements in a cube-sphere mesh. Each face of the cube has $120 \times 120$ surface elements, which is a frequent test case for HOMME on Titan. On Titan, HOMME obtains its best performance without threads; therefore, we run the MPI-only HOMME without threads. The strong scaling experiments are run from $10,800$ to $86,400$ processors.

As opposed to BlueGene/Q's contiguous allocations, Titan returns possibly sparse allocations. The default MPI rank ordering within an allocation is based on a Hilbert space filling curve that prioritizes the visit order based on the link bandwidths. For example, before jumping through slow Y links, it traverses whole a box in the dimension of $a \times 2 \times 4$, where the length $a$ of the box along the X dimension depends on the allocation and can be between 2 and 5. Therefore, the default MPI rank ordering preserves locality in the network, as long as locality is also preserved by the application. Since HOMME uses Hilbert SFC by default, it maps well to the default MPI rank ordering; therefore, there is only a little room for improvement from task mapping.

We compare the performance the proposed task mapping method to HOMME's default SFC method. We use three variants of our mapping methods:

● **Z2_1:** Our proposed mapping method with FZ ordering is used to map and partition the mesh elements to processors within a single step (equivalent to Z2 in Section 5.2).

● **Z2_2:** The partitioning algorithm used for mapping is slightly changed. The number of processors used in this experiment is not a power of 2. Because MJ recursively bisects the domain, processors within a node are bisected early in the partitioning process when the number of processors is not a power of 2. For example, when $10,800$ processors are used, overall 675 nodes are allocated. When bisecting these nodes equally, one of the nodes will be split in the first bisection. This will cause processors within the split node to have tasks that are likely to be far apart. In Z2_2, the bisection algorithm performs an uneven bisection, distributing the nodes/tasks in a way that prevents split nodes early in the hierarchy. The uneven bisection uses the largest prime divisor to determine weights for the two resulting parts. Since $10,800 = 2^4 \times 3^3 \times 5^2$, the largest prime divisor is 5. Rather than split the processors into two parts containing half the processors ($(2.5/5) \times 10,800 = 5400$) each, it creates one part with $6480 = 3/5 \times 10,800$ processors and one with $4320 = 2/5 \times 10,800$ processors. This method also employs an architecture-specific optimization, in which it scales the distances between the node coordinates based on the link bandwidths. The distances corresponding to links are scaled by $\frac{1}{bandwidth}$, so that nodes across links with higher bandwidth appear to be "closer" to each other than those across slower links.

● **Z2_3:** As with Z2_2, this method bisects based on the largest prime divisor. In addition, it creates boxes of dimension $2 \times 2 \times 8$ and transforms the $3D$ node coordinates to $6D$ node coordinates. Three of the coordinates are a node's coordinate within a box; the other three coordinates are the coordinates of the box. The coordinates are scaled based on the bandwidth as before; however, the coordinates corresponding to the box coordinates are scaled with larger

TABLE 2
HOMME communication time with various mapping strategies and transformations, normalized with respect to the communication time on $8192$ processors using HOMME's SFC.

| | Ref | | SFC+Z2 | | | | | | Z2 | | | | | |
| | SFC | Sphere | Sphere+E | Cube | Cube+E | 2DFace | 2DFace+E | Sphere | Sphere+E | Cube | Cube+E | 2DFace | 2DFace+E |
|---|---|---|---|---|---|---|---|---|---|---|---|---|---|
| 8192 | **1.00** (51.49 s) | 1.12 | 1.15 | 1.10 | 1.04 | 1.06 | 1.01 | **1.00** | 1.04 | 1.07 | 1.04 | 1.04 | 1.02 |
| 16384 | 0.64(33.14 s) | 0.58 | 0.58 | 0.58 | 0.53 | 0.61 | 0.59 | 0.57 | 0.56 | 0.56 | 0.53 | 0.51 | **0.51** |
| 32768 | 0.44(22.79 s) | 0.38 | 0.39 | 0.39 | 0.37 | 0.58 | 0.45 | 0.36 | 0.36 | 0.38 | 0.36 | **0.32** | 0.33 |

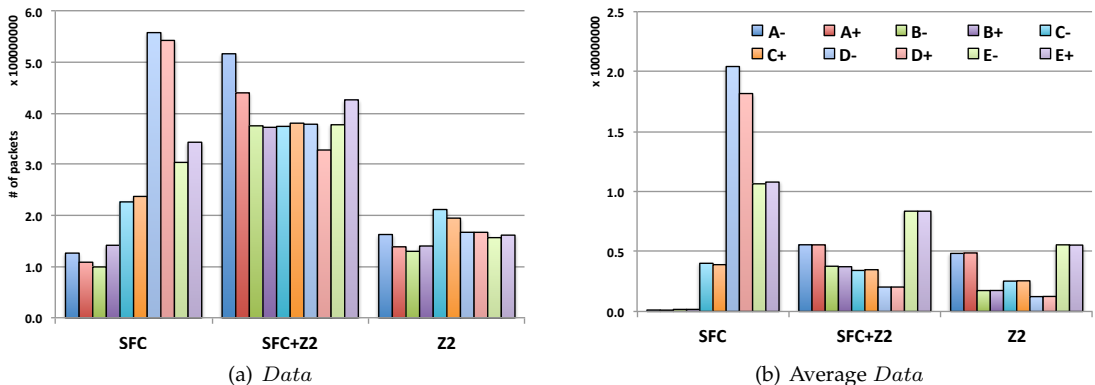

(a) $Data$          (b) Average $Data$

Fig. 9. $Data$ (Eqn. 5) and average $Data$ (Eqn. 4) on BG/Q links along different dimensions for Hybrid HOMME with 32K ranks.

weights to guide the partitioner to divide between boxes first, before dividing within boxes.

Figure 10 shows the communication time obtained with different mapping methods in HOMME. This experiment examines strong scaling, with the number of processors ranging from $10,800$ to $86,400$. We run three instances for each processor count, where each instance corresponds to a different allocation. Within each allocation, each mapping is executed twice. We experienced job interference due to external jobs on the system; therefore, for the sake of fair comparison, we include only those allocations where the execution time of the two mappings does not change significantly for all mappings run.

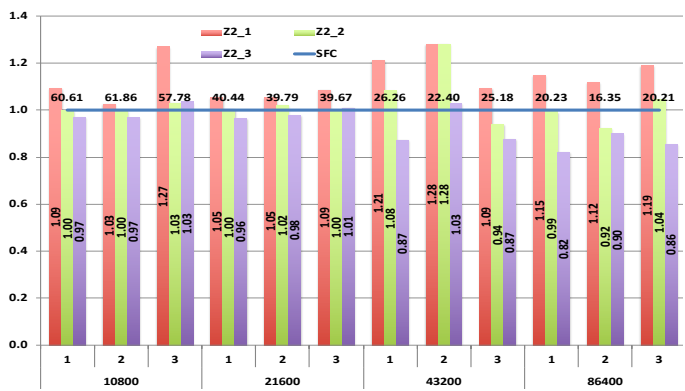

Fig. 10. HOMME's communication time using different mapping methods, normalized with respect to communication time using SFC. The numbers above the line are the communication time of SFC in seconds.

As seen in Figure 10, the improvements are minimal on HOMME on Titan. Z2_1 increases the communication time because nodes are assigned tasks that are far from each other. Z2_2 solves this problem partitioning based on the longest prime factor, which obtains similar performance to the default SFC. Only Z2_3 obtains reductions with respect to SFC, and these reductions increase as the number of processors increases. Z2_3 obtains up to $18\%$ reduction in the communication time of HOMME on $86,400$ processors.

Figure 11 shows the communication metrics for each of the allocations using Z2_3, normalized with respect to the metrics using SFC. In general, SFC obtains a good mapping that minimizes $WeightedHops$, $Latency$ (Eqn. 7), and total number of messages. In experiments with small numbers of processors, the Z2 mapping methods usually fail to improve most of these metrics, except for the total number of messages exchanged. As a result, their performance with low processor counts is usually close to SFC. On the other hand, on 86,400 processors, the partitions become identical (one task per part); therefore, changes in performance come only from the mapping methods. Z2_3 usually reduces $Latency$, while it increases the $WeightedHops$.

Figure 12 characterizes the network contention in each dimension on Titan for both SFC and Z2_3. Each result is normalized with respect to SFC in the X+ links. $Data(\mathcal{M})$ values do not consider the distinct bandwidths of the links, while $Latency$ values incorporate link bandwidths. SFC distributes $Data$ somewhat evenly, with most of the traffic going through Y- links. Thus, its $Latency$ is highest in the Y links. With its coordinate transformation methods, Z2_3 avoids traffic along the Y dimension and increases traffic along the X and Z dimensions. Although $Data$ is less evenly distributed by Z2_3, the traffic is well-distributed based on the bandwidth of the links. Therefore, $Latency$ along each dimension is more equal. This traffic distribution increases the utilization of the links, and reduces $Latency$ (and, as a result, the communication time), even though $WeightedHops$ increase by $25\%$. Because HOMME's messages are large, these bandwidth-based metrics are more important than latency-based ones.

In summary, when both applications and nodes employ similar ordering techniques, the room for improvement is likely to be small. However, performance can still be improved when application-specific requirements such as messages sizes and architecture specifications such as the non-uniform bandwidths are considered. In the next section, we study an application that uses a different task ordering than the underlying architecture.

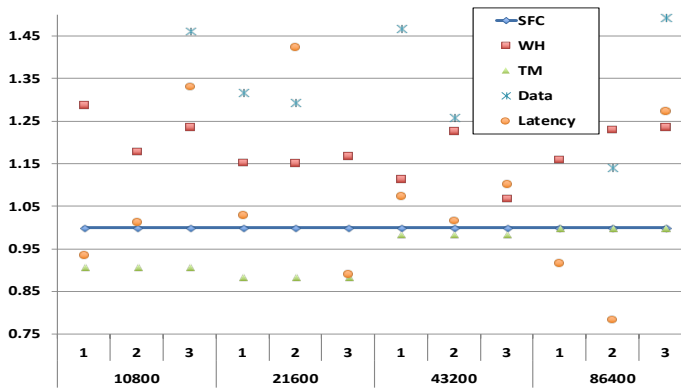

Fig. 11. Communication metrics in HOMME with Z2_3, normalized with respect to the metrics with SFC: WH ($WeightedHops$, Eqn. 3); TM (Total number of messages); $Data(\mathcal{M})$ (Eqn. 5); $Latency(\mathcal{M})$ (Eqn. 7).

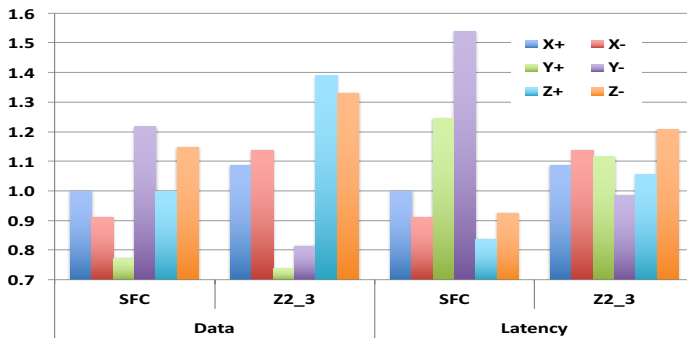

Fig. 12. With SFC and Z2_3, HOMME's $Data(\mathcal{M})$ and $Latency(\mathcal{M})$ in each network dimension (X+, X-, Y+, Y-, Z+, Z-), normalized by SFC X+

### 5.3.2 MiniGhost Task Mapping on Titan

MiniGhost [5] is a proxy application that implements a finite difference stencil and explicit time-stepping scheme across a three-dimensional uniform grid. Using a seven-point stencil, each task communicates with two neighbors along each dimension; tasks along geometry boundaries communicate with only their neighbors interior to the boundary (i.e., boundary conditions are non-periodic). Each task is assigned a subgrid of the 3D grid based on its task number. The number of tasks $tnum_x$, $tnum_y$, and $tnum_z$ in each dimension (with $tnum = (tnum_x)(tnum_y)(tnum_z)$) is specified by the user. Subgrids of the 3D grid are assigned to tasks by sweeping first in the $x$ direction, then the $y$ direction, and then the $z$ direction. Thus, task $i$ shares sub-grid boundaries (and, thus, requires communication) with tasks $i+1$ and $i-1$ to its east and west, respectively; with tasks $i + tnum_x$ and $i - tnum_x$ to its north and south; and with tasks $i + (tnum_x)(tnum_y)$ and $i - (tnum_x)(tnum_y)$ to its front and back. In the default MiniGhost configuration, task $i$ is performed by rank $i$. MiniGhost also provides an application-specific grouping of tasks for multicore nodes; on Titan, this Group method reorders tasks into $2 \times 2 \times 4$ blocks to better align tasks with the 16 cores in a node.

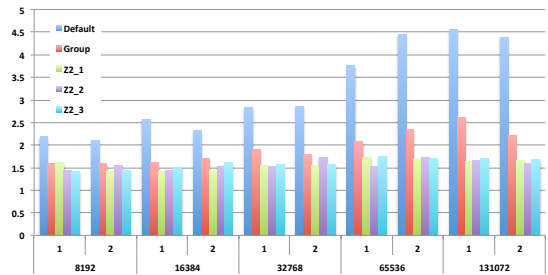

Fig. 13. Maximum communication time in MiniGhost (weak scaling)

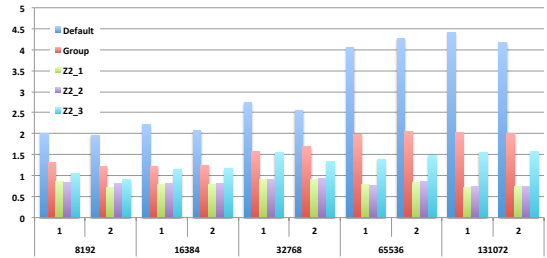

(a) $AverageHops$

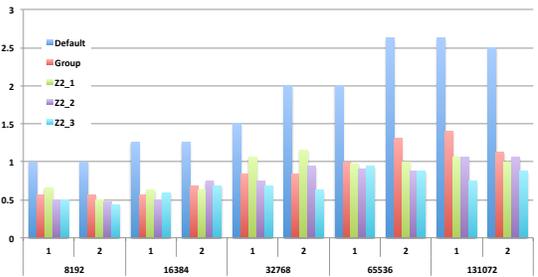

(b) $Latency(\mathcal{M})$

Fig. 14. $AverageHops$ and $Latency(\mathcal{M})$ in MiniGhost (weak scaling)

We ran weak scaling experiments using MiniGhost to evaluate the effect of mapping on communication time. We compared our geometric method with the MiniGhost's default task layout and its application-specific grouping. In these experiments, both the application connectivity and the network are three-dimensional. We used the same mapping variants Z2_1, Z2_2, and Z2_3 as before.

As shown in [4], the execution time of MiniGhost with its default mapping does not scale well in weak scaling tests. Our goal is to increase scalability by increasing locality of tasks within the allocation. We ran weak-scaling experiments with 8K–128K processors. Each task owned a $60 \times 60 \times 60$-cell subgrid; we ran the simulation for 20 timesteps with 40 variables per grid point. For each experiment, we obtained a node allocation of the requested size, and ran all mapping methods within that allocation. We repeated each experiment twice with different allocations.

Figure 13 shows the maximum communication time (across processors) for weak scaling experiments with MiniGhost. With MiniGhost's default mapping (Default), communication time increases dramatically as the number of processors is increased. The Group method controls the growth in communication costs, but consistent with results in [4], costs increase at the highest processor counts. Our geometric mapping methods provide the lowest communication costs, and, as desired for weak scaling, the com-munication costs remain nearly constant as the number of processors increases. Among the Z2 variants, Z2_1 and Z2_2 had similar performance. Their mappings differ only slightly, as the problem sizes are powers of two. On the other hand, Z2_3 obtains slightly slower performance.

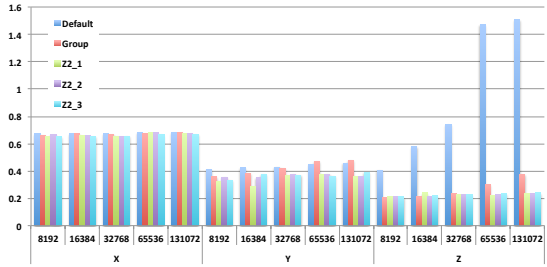

Fig. 15. Average communication time in each dimension in MiniGhost (weak scaling)

Figure 14 shows the metrics $AverageHops$ (Eqn. 2) and $Latency$ (Eqn. 7). With the default MiniGhost mapping, $AverageHops$, $Latency$ and communication time all follow the same upward trend as the number of processors increases. Since Group does not account for inter-node communication, its $AverageHops$ increase with the number of nodes. $AverageHops$ for the geometric mappings Z2_1 and Z2_2, however, remain nearly unchanged as we use more nodes, suggesting greater scalability using the geometric mappings. $Latency$ is also low with the geometric mappings, resulting in lower communication time. As in HOMME, Z2_3 obtains the lowest $Latency$ in most cases. However, its $AverageHops$ are up to twice as high as with Z2_1 and Z2_2. MiniGhost's messages are smaller (1 MB) than HOMME's; thus, reducing $Latency$ while doubling $AverageHops$ does not improve performance.

Overall, our geometric mapping methods reduced communication time by 35-64% relative to the default MiniGhost mapping, and 10-28% relative to the application-specific Group mapping. The largest reductions were seen at the highest processor counts, reflecting the importance of mapping as the number of cores in parallel computers increases.

## 6 Conclusion

We have proposed a new topology-aware task mapping method that uses multijagged geometric partitioning to reorder task and processor coordinates in a way that assigns communicating tasks to "nearby" processors. This method is designed for mesh and torus-based networks with contiguous and non-contiguous node allocations, such as IBM's BlueGene/Q and Cray's XK7. We also proposed several strategies (e.g., multiple rotations, coordinate shifting, new ordering schemes) that improve geometric mapping relative to a baseline geometric method. We compared our methods with the default mapping in two applications, as well as with application-specific mapping. Our geometric mapping reduced communication time up to 75% on 128K cores relative to the default mapping in the MiniGhost finite difference proxy application, and up to 31% on 32K ranks for the E3SM/HOMME climate modeling code. Our method can be applied to various applications and networks with heterogenous links by applying transformations to the coordinates to represent the application and network characteristics. Our implementation is open-source software in Zoltan2, available in Trilinos at https://github.com/trilinos/Trilinos.

We have also presented guidance about when task mapping is essential to reduce to communication time. We showed that applications obtain somewhat good mapping if their task placement is aligned with the network's MPI rank placement, while the mapping quality is low when the task placement schemes do not correspond. We showed that task mapping methods provide a portable way of obtaining good mappings for different network architectures. As future work, our mapping methods will be extended to accommodate dragonfly networks such as the Cray Aries network. We will investigate coordinate transformations to represent the hierarchies within the dragonfly networks.

## Acknowledgment

We thank Richard Barrett, Erik Boman, Jim Brandt, Ann Gentile, Torsten Hoefler, Vitus Leung, Stephen Olivier, Steve Plimpton, Christian Trott, and Courtenay Vaughan for helpful discussions. Sandia National Laboratories is a multimission laboratory managed and operated by National Technology and Engineering Solutions of Sandia LLC, a wholly owned subsidiary of Honeywell International Inc. for the U.S. Department of Energy's National Nuclear Security Administration under contract DE-NA0003525. The views expressed in the article do not necessarily represent the views of the U.S. Department of Energy or the United States Government.

# Appendix A

## Analysis of the Orderings

In this section, we study the number of hops induced by Z and Flipped-Z orderings. For simplicity, we assume that we have a one-to-one mapping of tasks to nodes, and that the number of tasks and nodes is a power of two, $2^n$. We assume a $pd$-dimensional mesh-topology processor network, and a $td$-dimensional tasks with a $td$-dimensional stencil-based communication pattern (i.e., each task communicates with $2 \times td$ neighbors). $n$ is divisible by both $td$ and $pd$. For the analysis, we assume that the cuts in the task- and processor-partitions are made in a consistent order. (This assumption may not apply with longest-dimension partitioning; we will comment about that case later.)

### A.1 Z Order

Z order is the default part numbering in our geometric partitioning methods. Using the example in Figure 3(a), we see the order of the cuts is $[gray, blue, red, green, pink, orange]$ We use 0-based reverse indices to represent these cuts; i.e., the indices of $orange$, $pink$, $green$, $red$, $blue$, and $gray$ are $0, 1, 2, 3, 4$, and $5$, respectively. Let $cuts_i$ be the sorted list of cut indices that is applied along dimension $i$. For example,

- $cuts_0 = cuts_x = [orange, green, blue] = [0, 2, 4]$.
- $cuts_1 = cuts_y = [pink, red, gray] = [1, 3, 5]$

Let $p$ be the binary representation of a part number that is assigned with Z ordering. And let $p^i$ denote the binary number that is obtained by filtering only the bits corresponding to $cuts_i$. The ordering along each dimension $i$ is affected only by $cuts_i$, and consecutive parts along dimension $i$ have consecutive $p^i$. For example, the bottom row of Figure 3(a) includes part numbers $\{0, 1, 4, 5, 16, 17, 20, 21\}$. Because $cuts_x = [0, 2, 4]$, we find

- $p = 0 =$ `000000`, $p^x = 000 = 0$
- $p = 1 =$ `000001`, $p^x = 001 = 1$
- $p = 4 =$ `000100`, $p^x = 010 = 2$
- $p = 5 =$ `000101`, $p^x = 011 = 3$
- $p = 16 =$ `010000`, $p^x = 100 = 4$
- $p = 17 =$ `010001`, $p^x = 101 = 5$
- $p = 20 =$ `010100`, $p^x = 110 = 6$
- $p = 21 =$ `010101`, $p^x = 111 = 7$

As a result, for two neighboring parts $p_1$ and $p_2$ along dimension $i$, $p_1^i = p_2^i + 1$. Further, when $p_1^i$ has a pattern `xxx...xxx0111...111`, where there are $j$ consecutive 1s starting from the least significant bit (Bit 0), $p_2^i$ has a pattern `xxx...xxx1000...000`. $p_1^i$ and $p_2^i$ differ in their $j+1$ least significant bits.

Neighbors differ in the least significant $j + 1$ bits their $p^i$ are separated by the cut with index $j$ in dimension $i$. For example, in Figure 3(a), the blue cut has index $j = 2$ along the $x$ dimension. It separates neighbors

- $p_1 = 5 =$ `000101`, $p_1^x = 011$
- $p_2 = 16 =$ `010000`, $p_2^x = 100$

on the first row, which differ in $j + 1 = 3$ red bits.

In 1D, the number of neighbors that are separated by a cut with index $j$ along dimension $i$ is

$$NN1D_i(j) = \frac{2^{|cuts_i|}}{2^j}. \tag{8}$$

Since $NN1D_i(j)$ is replicated along all other dimensions, there are $\frac{2^n}{2^{|cuts_i|}}$ replications. Thus, the overall number of neighbors that are separated by the cut with index $j \in cuts_i$ along dimension $i$ is

$$NN_i(j) = 2^{n-j} \tag{9}$$

We analyze the number of hops for the messages associated with each cut index $j$ along a task dimension $td_i$. Let

- $cuts_{td_i} = [x | x = i + td \times j \text{ and } j \in [0, \frac{n}{td}]]$
- $cuts_{pd_i} = [x | x = i + pd \times j \text{ and } j \in [0, \frac{n}{pd}]]$

for $td_i = 0, 1, \ldots, td - 1$, and $pd_i = 0, 1, \ldots, pd - 1$.

Figure 16(a) shows a four-dimensional ($td = 4$) partition and numbering of two tasks that are neighbors in task dimension $td_3$. Each dimension's partition is shown with a different color; the bits with red, blue, yellow and green colors correspond to dimensions $td_0$, $td_1$, $td_2$, and $td_3$, respectively. We have $cuts_{td_3} = [3, 7, 11, 15, 19, \ldots]$, and we consider the neighborhood of two consecutive numbers that are separated by the cut $j = 4$ along dimension $td_3$. Such neighbors differ in the least significant $j + 1$ bits of $p^{td_3}$.

Since tasks are mapped to processors that have the same part numbers, we need to find the number of hops between the processors that share these tasks' part numbers. Figure 16(b) shows an example of the processor partition and the bits for the case where $pd = 1$. In this case, the number of hops between the neighbors is the difference between the part numbers $(p_2^{pd_0} - p_1^{pd_0}) = 2^{19} - 2^{15} - 2^{11} - 2^7 - 2^3$.

Similarly, Figure 16(c) shows processor part assignments for $pd = 2$. In this case, the number of hops along dimension $pd_0$ (red bits) is $(p_2^{pd_0} - p_1^{pd_0}) = 0$, while the number of hops along $pd_1$ (blue bits) is $(p_2^{pd_1} - p_1^{pd_1}) = 2^9 - 2^7 - 2^5 - 2^3 - 2^1$.

The number of hops for $pd = 3$, $pd = 4$ and $pd = 8$ is explained in Figure 16. In general, the number of hops for neighbors that are separated by cut $j$ along $td_i$ (neighbors that differ in the $j + 1$ least significant bit of $p^{td_i}$) is

$$NHZ_{td_i}(j) = 2^{\lfloor \frac{td \times j + i}{pd} \rfloor} +$$
$$\sum_{k=0}^{j-1} 2^{\lfloor \frac{td \times k + i}{pd} \rfloor} \quad sign((td \times k + i) \pmod{pd},$$
$$(td \times j + i) \pmod{pd}) \tag{10}$$

where $sign(a, b) = -1$ if $a = b$; 1, otherwise. That is, $b = (td \times j + i) \pmod{pd}$ is the processor dimension of the most significant differing bit in the node partition, and $\lfloor \frac{td \times j + i}{pd} \rfloor$ is the position of that bit in $p^b$. In Z ordering, differing bits that are in the same dimension with the most significant bit reduce the number of hops, while the differing bits in other dimensions increase the number of hops. Thus, if $pd$ and $td$ are similar and somewhat structured, Z ordering is likely to reduce the number of hops. We simplify Equation 10 for further comparisons:

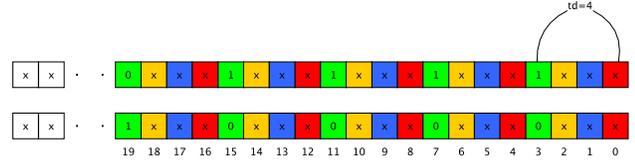

(a) Task partition with $td = 4$. These tasks are neighbors along task dimension $td_3$.

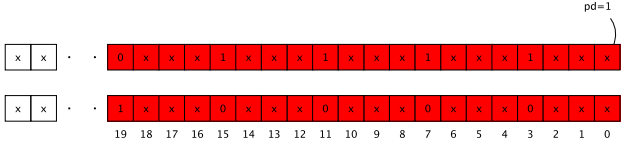

(b) Processor partition with $pd = 1$. The number of hops between the two consecutive task numbers in a one-dimensional processor network is the difference of the two numbers:
$p_2^{pd_0} - p_1^{pd_0} = 2^{19} - 2^{15} - 2^{11} - 2^7 - 2^3$.

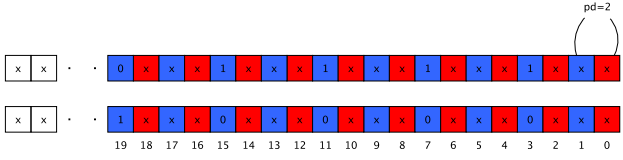

(c) Processor partition with $pd = 2$. The number of hops along processor dimension $pd_0$ (red) is zero, since all red bits are identical ($p_2^{pd_0} - p_1^{pd_0} = 0$). The number of hops along $pd_1$ (blue) is the difference of the blue bits: $p_2^{pd_1} - p_1^{pd_1} = 2^9 - 2^7 - 2^5 - 2^3 - 2^1$.

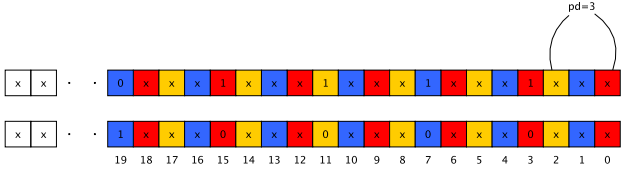

(d) Processor partition with $pd = 3$. The number of hops along processor dimensions $pd_0$ (red), $pd_1$ (blue), and $pd_2$ (yellow) are $2^5 + 2^1$, $2^6 - 2^2$, and $2^3$, respectively.

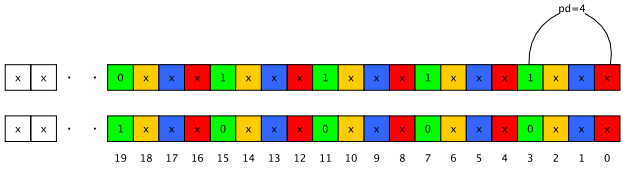

(e) Processor partition with $pd = 4$. Because $pd = td$, the task and processor partitions align. Thus, the number of hops is zero for $pd_0$ (red), $pd_1$ (blue), and $pd_2$ (yellow), and one for $pd_3$ (green).

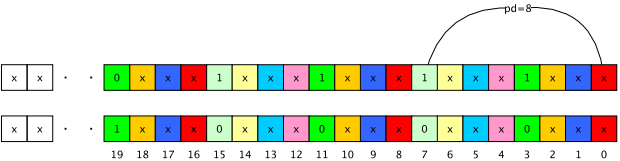

(f) Processor partition with $pd = 8$. The bit patterns for dimensions $pd_0$, $pd_1$, $pd_2$, $pd_4$, $pd_5$, and $pd_6$ are identical, so there are zero hops in these dimensions. The number of hops is $2^2 - 2^1 - 2^0$ along $pd_3$ (green), and $2^1 + 2^0$ along $pd_7$ (light green).

Fig. 16. Partitioning of two consecutive part numbers in Z order with task dimension $td = 4$ and varying processor dimensions $pd = 1, 2, 3, 4, 8$. The bit patterns of two part numbers that are consecutive in Z-order are shown in each subfigure; an "x" in a bit position indicates that the two part numbers share the same value in that position. The partitioning of the bits to determine Z-order position in each dimension is indicated with colors. The number of hops in each dimension is calculated from the partitioning of the bit pattern.

| Decimal | Binary | FZ | Gray Code | Decimal | Binary | FZ | Gray Code |
|---|---|---|---|---|---|---|---|
| 0 | 00000 | 0 | 00000 | 16 | 10000 | 24 | 11000 |
| 1 | 00001 | 1 | 00001 | 17 | 10001 | 25 | 11001 |
| 2 | 00010 | 3 | 00011 | 18 | 10010 | 27 | 11011 |
| 3 | 00011 | 2 | 00010 | 19 | 10011 | 26 | 11010 |
| 4 | 00100 | 6 | 00110 | 20 | 10100 | 30 | 11110 |
| 5 | 00101 | 7 | 00111 | 21 | 10101 | 31 | 11111 |
| 6 | 00110 | 5 | 00101 | 22 | 10110 | 29 | 11101 |
| 7 | 00111 | 4 | 00100 | 23 | 10111 | 28 | 11100 |
| 8 | 01000 | 12 | 01100 | 24 | 11000 | 20 | 10100 |
| 9 | 01001 | 13 | 01101 | 25 | 11001 | 21 | 10101 |
| 10 | 01010 | 15 | 01111 | 26 | 11010 | 23 | 10111 |
| 11 | 01011 | 14 | 01110 | 27 | 11011 | 22 | 10110 |
| 12 | 01100 | 10 | 01010 | 28 | 11100 | 18 | 10010 |
| 13 | 01101 | 11 | 01011 | 29 | 11101 | 19 | 10011 |
| 14 | 01110 | 9 | 01001 | 30 | 11110 | 17 | 10001 |
| 15 | 01111 | 8 | 01000 | 31 | 11111 | 17 | 10000 |

$$NHZ_{td_i}(j) \begin{cases} = 1, & \text{if } td = pd; \\ < 2^{\lfloor \frac{td \times j + i}{pd} \rfloor}, & \text{if } td \ (\text{mod } pd) = 0; \\ > 2^{\lfloor \frac{td \times j + i}{pd} \rfloor}, & \text{otherwise.} \end{cases} \quad (11)$$

That is, if the two most significant bits are in the same dimension with respect to the processor partition dimensions, the number of hops is strictly less than $2^{\lfloor \frac{td \times j + i}{pd} \rfloor}$; otherwise, the number of hops is greater.

### A.2 Flipped-Z Order

Flipped-Z (FZ) order performs ordering that is somewhat in between Z-order and Gray order (G). G flips all coordinates of the larger half after each bisection, while FZ flips only those coordinates corresponding to the partitioning dimension. When applied to 1D data, FZ returns the same order as G, as shown in Table 3. Consecutive indices differ by only a single bit in Gray order. Two observations follow.

- Two neighbor parts in FZ that are separated with a cut with index $j$ differ only in bit $j$ of their Gray-code representation. For example, the FZ parts 0 and 1 are separated by the cut with index $j = 0$; they differ only in the least significant bit. Similarly, neighboring FZ-order parts 8 (Gray-code 01000) and 24 (11000) are separated by the cut with index $j = 4$; they differ only in bit 4.

- If two neighbors differ in bit $j$ of their Gray-code value, all the bits up to and including bit $j - 2$ are 0, and bit $j - 1$ is 1. For FZ-order tasks 8 and 24, bits 0, 1, and 2 are zero bits, and bit $j - 1 = 3$ is 1 in both.

As with Z, the FZ ordering along dimension $i$ is affected only by $cuts_i$. FZ induces a Gray order on $p^i$, the bits corresponding to $cuts_i$. For example, in the bottom row of Figure 3(d), the parts are $\{0, 1, 5, 4, 20, 21, 17, 16\}$; this order corresponds to Gray ordering of the red bits $p^x$.

- $p = 0 = \ \textcolor{red}{000}00$, $p^x = 000$
- $p = 1 = \ \textcolor{red}{000}01$, $p^x = 001$
- $p = 5 = \ \textcolor{red}{001}01$, $p^x = 011$
- $p = 4 = \ \textcolor{red}{001}00$, $p^x = 010$
- $p = 20 = \textcolor{red}{101}00$, $p^x = 110$
- $p = 21 = \textcolor{red}{101}01$, $p^x = 111$

- $p = 17 =$ `010001`, $p^x = $ `101`
- $p = 16 =$ `010000`, $p^x = $ `100`

Figure 17(a) shows a 4-dimensional ($td = 4$) partition and numbering of two consecutive tasks in dimension $td_3$ with FZ. Each dimension's partition is shown with a different color. The bits with red, blue, yellow and green colors correspond to dimensions $td_0$, $td_1$, $td_2$, and $td_3$, respectively. We consider the neighborhood of two consecutive numbers that are separated by the cut with index $j$ along dimension $td_3$; such neighbors differ in bit $j = 4$ in the figure. Figure 17(b) shows the processor partition and the distribution of the bits for $pd = 1$. In this case, the number of hops between neighboring tasks is the difference between the decimal task numbers corresponding to the Gray code as in Table 3. Because of Gray code's properties, the distance between two Gray-code numbers depends on the values of the x's in Figure 17(a); that is, the distance is different for each pair of neighbors. For example, Figure 18 contains a simple case with the Gray code for eight parts. Looking at Gray-code pairs that differ only in bit 2, we see that the number of hops between them is not a simple difference of the Gray-code bits. The actual distance between such pairs depends on the values of the lower-order bits that they share.

However, it is possible to calculate the *average* number of hops per message in FZ using the Gray-code values. To calculate the average number of hops, we define the most significant free bit, $MSFB$, to be the most significant bit with "x" values. In Figure 17(b), consecutive pairs that differ in bit 19 with respect to $pd_0$ are shown; $MSFB$ is bit 18. In Gray code, half of such pairs have 0 for their $18^{th}$ bit, and half have 1. Therefore, for each pair in the form

- $p_1 = $ `01xx`
- $p_2 = $ `11xx`

there is another pair

- $p_3 = $ `00xx`
- $p_4 = $ `10xx`

Because of the properties of Gray code, if

- $p_3 = y$ in decimal, then
- $p_4 = 2^{20} - y - 1$ in decimal.

Similarly,

- $p_1 = 2^{19} - y - 1$ in decimal, and
- $p_2 = 2^{20} - 2^{19} + y$ in decimal.

The sum of the number of hops between these two pairs is

- $[(2^{20} - y - 1) - y] + [(2^{20} - 2^{19} + y) - (2^{19} - y - 1)] = 2^{20}$.

For each set of pairs ($p_1$, $p_2$) and ($p_3$, $p_4$), the sum of the hops is $2^{20}$, making the average number of hops $2^{19}$ for pairs differing in bit 19. For example,

- $p_1 = $ `01111000000000000000` (Gray)
  327680 (decimal)
- $p_2 = $ `11111000000000000000` (Gray)
  720895 (decimal)
- $p_3 = $ `00111000000000000000` (Gray)
  196607 (decimal)

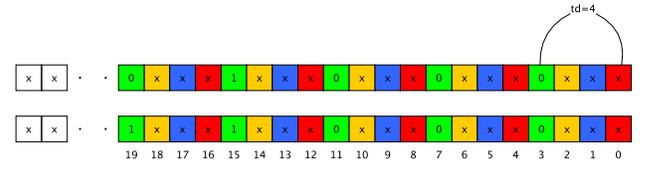

(a) Task partition with $td = 4$. These tasks are neighbors along task dimension $td_3$.

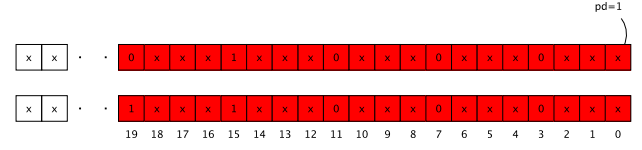

(b) Processor partition with $pd = 1$. The average number of hops between parts differing in bit 19 along processor dimension $pd_0$ is $2^{19}$. (See discussion in the text.)

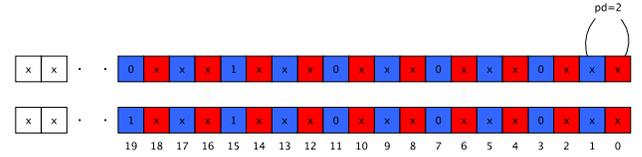

(c) Processor partition with $pd = 2$. The average number of hops along processor dimension $pd_1$ for parts differing in bit 9 of $p^{pd_1}$ (blue) is $2^9$; the number of hops along $pd_0$ is zero.

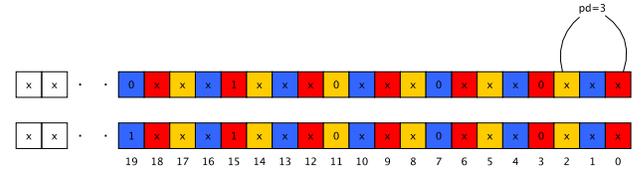

(d) Processor partition with $pd = 3$. Along processor dimensions $pd_0$ (red) and $pd_2$ (yellow), there are zero hops between the parts. The average number of hops along $pd_1$ (blue) for parts differing in bit 6 of $p^{pd_1}$ is $2^6$.

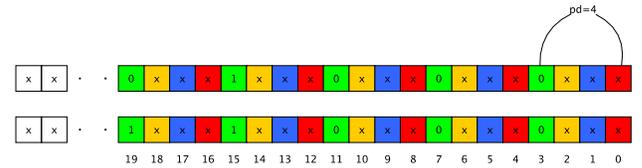

(e) Processor partition with $pd = 4$. Because $pd = td$, the task and processor partitions align. Thus, the average number of hops is zero for $pd_0$ (red), $pd_1$ (blue), and $pd_2$ (yellow), and one for $pd_3$ (green).

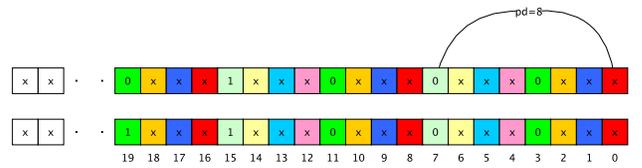

(f) Processor partitioning with $pd = 8$. The number of hops is zero along processor dimensions $pd_0$, $pd_1$, $pd_2$, $pd_4$, $pd_5$, $pd_6$ and $pd_7$. The average number of hops along $pd_3$ (green) is $2^3 - 1 = 7$. (See the discussion of conflict bits in the text.)

Fig. 17. Partitioning of two consecutive part numbers in FZ order with task dimension $td = 4$ and varying processor dimensions $pd = 1, 2, 3, 4, 8$. Gray-code bit patterns of two part numbers that are consecutive in task dimension $td_3$ are shown in each subfigure; an "x" in a bit position indicates that the two part numbers share the same value in that position. The partitioning of the bits to determine FZ-order position in each dimension is indicated with colors. The average number of hops in each dimension is calculated from the partitioning of the bit pattern.

Fig. 18. A simple eight-part example showing how, with FZ order, the values of bits 0 and 1 in each Gray-code part number affect the number of hops between pairs of parts that differ only in bit 2. In the example, the *average* number of hops between pairs differing only in bit 2 is four.

- $p_4 =$ `10111000000000000000` (Gray)
  851968 (decimal)

The sum of the number of hops between these pairs is $720895 - 327680 + 851968 - 196607 = 2^{20}$.

In Figure 17(c) with $pd = 2$, the mapping has $2^9$ hops on average along $pd_1$ (blue), and 0 along $pd_0$ (red) for the task pairs in Figure 17(a). Likewise, in Figure 17(d) with $pd = 3$, the mapping has $2^6$ hops on average along $pd_1$ (blue) and zero hops along $pd_0$ (red) and $pd_2$ (yellow). In fact, since the neighboring parts differ by only a single bit, they always have hops along only a single dimension.

In Figure 17(e), $pd = 4$, so the task and processor partitions are exactly the same. In this case, there are no free bits along $pd_3$ ($MSFB = -1$). Thus, the average number of hops along $pd_3$ is $2^0 = 1$.

For $pd = 8$ in Figure 17(f), $pd_3$ (green) has the differing bit; its neighboring tasks along $pd_3$ are `000` and `100`. In this example, there are no free bits with "x" values in $pd_3$. Instead, the second most significant bit is 0. This is a conflict case for FZ, causing the consecutive tasks to have the maximum number of hops between them; i.e, `000` and `100` are the furthest tasks from each other considering only 3 bits. If the second most significant bits were "1" instead of "0" (`010` and `110`), they would have only a single hop between them. In general, if there is such a conflicting bit at position $CB$, the number of hops is increase by $2^{CB+2} - 1$. In more general way, if $MSFB < CB$, the average number of hops is $2^{CB+2} - 2^{MSFB+1}$. In this example, $CB = 1$, and $MSFB = -1$ (does not exist), so the number of hops between `000` and `100` is 7.

The average number of hops for FZ along dimension $td_i$ for tasks that differ in bit $j$ is then

$$NHF_{td_i}(j) = \begin{cases} 1, & \text{if } td = pd; \\ 2^{\lfloor \frac{td \times j + i}{pd} \rfloor + 1} - 1, & \text{else if } pd \,(\text{mod } td) = 0; \\ 2^{\lfloor \frac{td \times j + i}{pd} \rfloor}, & \text{otherwise.} \end{cases}$$ (12)

The first case is trivial. The second case results in conflict bits, with $CB = j - 1$ and $MSFB = -1$. The third case is the general case, in which $td$ and $pd$ are not factors of each other, which results in $MSFB = j - 1$.

Comparing Equations 11 and 12, we see

- $NHF_{td_i}(j) = NHZ_{td_i}(j)$, if $td = pd$. The orderings are equivalent when $td = pd$ and the cuts have consistent order.
- $NHF_{td_i}(j) > NHZ_{td_i}(j)$, if $td \,(\text{mod } pd) = 0$. Z ordering is likely to have better mapping when $td$-dimensional tasks are mapped to $pd = td \times k$ processor dimensions. Yet even for this case, Z ordering might suffer if the cuts do not have consistent ordering.
- $NHF_{td_i}(j) < NHZ_{td_i}(j)$, if $pd \,(\text{mod } td) \neq 0$. If $td$ and $pd$ are not factors of each other, FZ is likely to produce better mappings.
- The case with $pd \,(\text{mod } td) = 0$ is the worst case for FZ. Yet, below we show that $NHF_{td_i}(j) < NHZ_{td_i}(j)$ for $pd \,(\text{mod } td) = 0$.

Throughout the analysis, we assumed that consistent cut orderings are used in both the task and processor partitions. However, with longest-dimension partitioning, partitioning can be done in arbitrary dimensions; e.g., two consecutive cuts can be along the same dimension as in Figure 2. The quality of the Z ordering increases as with the overlap of the partitioning patterns of the tasks and processors. We expect that such pattern overlaps are much smaller in mappings with longest dimension partitioning. Therefore, we expect FZ to be better than Z. In addition, our analysis assumes that both task- and processor-networks are meshes. Because of the cyclic properties of Gray code, we expect FZ orderings to be even better on torus structures, as is demonstrated in the experiments of Section 5.

### A.3 Number of Hops when $pd \,(\text{mod } td) = 0$

We analyze the special case in which $pd \neq td$ and $(pd \,(\text{mod } td)) = 0$. Let $m = \frac{pd}{td}$. Equation 12 then becomes

$$NHF_{td_i}(j) = 2^{\lfloor \frac{j}{m} \rfloor + 1} - 1.$$ (13)

In computing $NHZ_{td_i}(j)$ for this case, cut indices $k$ on the processor side (i.e., in $cuts_{pd_{im}}$, $cuts_{pd_{im+1}}, \ldots, cuts_{pd_{(i+1)m)-1}}$) can be calculated as $k = \lfloor \frac{j}{m} \rfloor$, where $j$ is a cut index on the task side (i.e., in $cuts_{td_i}$). There are $m$ cut indices that produce the same value for $k$. Thus, in Equation 10, the sign function returns $-1$ once every $m$ indices, while it returns 1 for the other $m-1$ indices. Thus, we can rewrite Equation 10 as

$$NHZ_{td_i}(j)$$
$$= 2^{\lfloor \frac{j}{m} \rfloor}(j \,(\text{mod } m) + 1) + (m - 2) \sum_{k=0}^{\lfloor \frac{j}{m} \rfloor - 1} 2^k$$
$$= 2^{\lfloor \frac{j}{m} \rfloor}(j \,(\text{mod } m) + 1) + (m - 2)(2^{\lfloor \frac{j}{m} \rfloor} - 1)$$
$$= 2^{\lfloor \frac{j}{m} \rfloor}(j \,(\text{mod } m)) + (m - 1)(2^{\lfloor \frac{j}{m} \rfloor}) + 2 - m$$ (14)

For $m \geq 3$, $NHZ_{td_i}(j)$ in Equation 14 is always larger than $NHF_{td_i}(j)$ in Equation 13; thus, FZ order is preferred.

However, for $m = 2$, the number of hops with respect to cut $j$ in Z ordering is

$$NHZ_{td_i}(j) = \begin{cases} 2^{\frac{j}{2}}, & \text{if j is even;} \\ 2^{\frac{j-1}{2}+1}, & \text{otherwise.} \end{cases}$$ (15)

To simplify the presentation, let $C = |cuts_{td_i}|$. From Equation 8, we have $NN1D_{td_i}(j) = 2^{C-j}$ messages across cut $j$. Therefore, using Z, the number of hops for all messages across cut $j$ is

$$NeighborsHopsZ_{td_i}(j) = NN1D_{td_i}(j) \times NHZ_{td_i}(j)$$
$$= \begin{cases} 2^{C-\frac{j}{2}}, & \text{if j is even;} \\ 2^{C-\frac{j-1}{2}}, & \text{otherwise.} \end{cases} \quad (16)$$

We compute the total number of hops with respect to $td_i$ and all cuts by summing Equation 16 over all cuts $j \in cuts_{td_i}$. First, assume that $C$ is even. The total number of hops for the even cuts $(j = 2k)$ is $\sum_{k=0}^{\frac{C}{2}-1} 2^{C-k}$; the number of hops for the odd bits $(j = 2k+1)$ is also $\sum_{k=0}^{\frac{C}{2}-1} 2^{C-k}$.

$$TotalHopsZ_{td_i} = 2 \sum_{k=0}^{\frac{C}{2}-1} 2^{C-k}$$
$$= 2(2^C) \sum_{k=0}^{\frac{C}{2}-1} 2^{-k}$$
$$= 2(2^C)[\frac{(\frac{1}{2})^{\frac{C}{2}}-1}{\frac{1}{2}-1}]$$
$$= 2^2(2^C)(1-2^{\frac{-C}{2}})$$
$$= 2^{C+2}-4(2^{\frac{C}{2}}). \quad (17)$$

If $C$ is odd, there is one extra term in the summation for the even cuts, which becomes $\sum_{k=0}^{\frac{C-1}{2}} 2^{C-k}$; the summation for the odd cuts becomes $\sum_{k=0}^{\frac{C-1}{2}-1} 2^{C-k}$. Then,

$$TotalHopsZ_{td_i} = \sum_{k=0}^{\frac{C-1}{2}} 2^{C-k} + \sum_{k=0}^{\frac{C-1}{2}-1} 2^{C-k}$$
$$= 2^{C-\frac{C-1}{2}} + 2 \sum_{k=0}^{\frac{C-1}{2}-1} 2^{C-k}$$
$$= 2^{\frac{C+1}{2}} + 2(2^C) \sum_{k=0}^{\frac{C-1}{2}-1} 2^{-k}$$
$$= 2^{\frac{C+1}{2}} + 2(2^C)[\frac{(\frac{1}{2})^{\frac{C-1}{2}}-1}{\frac{1}{2}-1}]$$
$$= 2^{\frac{C+1}{2}} + 2^2(2^C)(1-2^{\frac{-C+1}{2}})$$
$$= 2^{\frac{C+1}{2}} + 2^{C+2} - 4(2^{\frac{C+1}{2}})$$
$$= 2^{C+2} - 3(2^{\frac{C+1}{2}}). \quad (18)$$

Therefore,

$$TotalHopsZ_{td_i} = \begin{cases} 2^{C+2}-4(2^{\frac{C}{2}}), & \text{if C is even;} \\ 2^{C+2}-3(2^{\frac{C+1}{2}}), & \text{otherwise.} \end{cases} \quad (19)$$

For Flipped Z order, Equation 13 with $m = 2$ becomes

$$NHF_{td_i}(j) = 2^{\lfloor \frac{j}{2} \rfloor + 1} - 1.$$

The number of hops for all messages across cut index $j$ is

$$NeighborsHopsF_{td_i}(j) = NN1D_{td_i}(j) \times NHF_{td_i}(j)$$
$$= (2^{C-j})(2^{\lfloor \frac{j}{2} \rfloor + 1} - 1). \quad (20)$$

Summing across all cuts $j \in cuts_{td_i}$, we find $TotalHopsF_{td_i}$, the total number of hops for all messages across all cuts $j \in cuts_{td_i}$. When $C$ is even, the number of hops for the even cuts $(j = 2k)$ is $\sum_{k=0}^{\frac{C}{2}-1} (2^{C-2k})(2^{k+1}-1)$, and the number of hops for odd cuts $(j = 2k+1)$ is $\sum_{k=0}^{\frac{C}{2}-1} (2^{C-2k-1})(2^{k+1}-1)$. The sum of the number of hops for the even and odd cuts is then

$$TotalHopsF_{td_i} = \sum_{k=0}^{\frac{C}{2}-1} 3(2^{C-2k-1})(2^{k+1}-1)$$
$$= 3(2^C) \sum_{k=0}^{\frac{C}{2}-1} (2^{-k}-2^{-2k-1})$$
$$= 3(2^C) \sum_{k=0}^{\frac{C}{2}-1} (2^{-k}-2^{-2k-1})$$
$$= 3(2^C)[\frac{(\frac{1}{2})^{\frac{C}{2}}-1}{\frac{1}{2}-1} - \frac{1}{2}(\frac{(\frac{1}{4})^{\frac{C}{2}}-1}{\frac{1}{4}-1})]$$
$$= 3(2^C)[(2-2^{\frac{-C}{2}+1} + \frac{2}{3}(2^{-C}-1)]$$
$$= 2^C[6-3(2^{\frac{-C}{2}+1}) + 2^{-C+1} - 2]$$
$$= 4(2^C) - 3(2^{\frac{C}{2}+1}) + 2$$
$$= 2^{C+2} - 6(2^{\frac{C}{2}}) + 2. \quad (21)$$

When $C$ is odd, the number of hops relative to even cuts $(j = 2k)$ is $\sum_{k=0}^{\frac{C-1}{2}} (2^{C-2k})(2^{k+1}-1)$, and relative to odd cuts $(j = 2k+1)$ is $\sum_{k=0}^{\frac{C-1}{2}-1} (2^{C-2k-1})(2^{k+1}-1)$. The total number of hops relative to all cuts is then

$$TotalHopsF_{td_i} = \sum_{k=0}^{\frac{C-1}{2}} (2^{C-2k})(2^{k+1}-1)+$$

$$\sum_{k=0}^{\frac{C-1}{2}-1} (2^{C-2k-1})(2^{k+1}-1)$$

$$= 2^{\frac{C+3}{2}} - 2 + \sum_{k=0}^{\frac{C-1}{2}-1} 3(2^{k+1}-1)(2^{C-2k-1})$$

$$= 2^{\frac{C+3}{2}} - 2 + 3(2^C) \sum_{k=0}^{\frac{C-1}{2}-1} (2^{-k} - 2^{-2k-1})$$

$$= 2^{\frac{C+3}{2}} - 2 + 3(2^C)[\frac{(\frac{1}{2})^{\frac{C-1}{2}}-1}{\frac{1}{2}-1} - \frac{1}{2}(\frac{(\frac{1}{4})^{\frac{C-1}{2}}-1}{\frac{1}{4}-1})]$$

$$= 2^{\frac{C+3}{2}} - 2 + 3(2^C)[2 - 2(2^{\frac{-C+1}{2}}) + \frac{2}{3}(2^{1-C}-1)]$$

$$= 2^{\frac{C+3}{2}} - 2 + 2^C[6 - 6(2^{\frac{-C+1}{2}}) + 2^{2-C} - 2]$$

$$= 2^{\frac{C+3}{2}} - 2 + 4(2^C) - 6(2^{\frac{C+1}{2}}) + 2^2$$

$$= 2^{\frac{C+3}{2}} - 2 + 2^{C+2} - 6(2^{\frac{C+1}{2}}) + 4$$

$$= 2^{C+2} - 4(2^{\frac{C+1}{2}}) + 2. \tag{22}$$

Therefore, the overall number of hops is

$$TotalHopsF_{td_i} = \begin{cases} 2^{C+2} - 6(2^{\frac{C}{2}}) + 2, & \text{if C is even;} \\ 2^{C+2} - 4(2^{\frac{C+1}{2}}) + 2, & \text{otherwise.} \end{cases} \tag{23}$$

Comparing the number of hops with FZ in Equation 23 with the number for Z in Equation 19, we can see that FZ obtains fewer hops in this case.